\pdfoutput=1
\documentclass[pdflatex,sn-basic]{sn-jnl}

\usepackage{graphicx}
\usepackage{multirow}
\usepackage{amsmath,amssymb,amsfonts}
\usepackage{booktabs}
\usepackage{algorithm}
\usepackage{algorithmicx}
\usepackage{algpseudocode}
\usepackage{longtable}
\usepackage{textcomp}
\usepackage{manyfoot}
\usepackage{float}
\usepackage{url}

\theoremstyle{definition}
\newtheorem{definition}{Definition}

\usepackage{pdfpages}

\begin{document}

\title[Diverse Minds, Divided Networks?]{Diverse Minds, Divided Networks? Personality
Composition, Polarization, and Collective Intelligence in LLM-Based Social Simulations}

\author*[1]{\fnm{Raad} \sur{Bin Tareaf}}\email{raad.bintareaf@german-uds.de}

\affil*[1]{\orgname{German University of Digital Science},
  \orgaddress{\city{Potsdam}, \country{Germany}}}

\abstract{Simulated societies of large language model agents are used to study online
polarization, and separately to study collective intelligence, but the two are rarely
measured in the same system. It is therefore difficult to say whether a society's personality
composition shapes both, or whether reducing polarization costs collective competence. We
present TraitMix, an experimental design in which the Big Five composition of a simulated
social network, both trait levels and trait heterogeneity, is a controlled experimental
variable, and in which polarization and collective performance are measured in the same runs.
Across 991 simulations of hundred-agent societies, spanning six contested topics and six
language models, trait heterogeneity has the largest measured effects, acting in opposite
directions on two faces of polarization: varied societies hold more dispersed opinions while
being less segregated into camps, so homogeneous societies are not moderate but consensual
echo chambers. Trait effects are not additive, as Agreeableness determines the sign of
Openness, an interaction that replicates across models although the primary model's estimate
is influence-driven. Contrary to the trade-off the study was designed to measure, no
polarization measure predicts poorer collective performance, and cross-cutting interaction is
the only one of four whose association with collective accuracy survives partialling on the
aggregation identity. We report ablations removing two potential measurement circularities,
an induction gate applied to every model, and the measures that failed them.}


\maketitle

\section{Introduction}\label{sec:intro}

Two questions about online societies are usually asked separately. Why do they split into
hostile camps? And when can a crowd think better than the individuals in it? Both have been
studied for decades, and both have recently been taken up again using large language models
as agents, which allow simulated populations to talk to one another in natural language
rather than exchange numbers.

The two literatures have not met. Studies of polarization in language-model societies
reproduce echo chambers, opinion clustering and human-like polarization
\cite{chuang2024simulating, wang2025decoding, piao2025polarization}, but treat the agents as
interchangeable: where Big Five personality traits appear at all, they are assigned as
background detail rather than manipulated, and no group task is measured. Studies of
collective intelligence in language-model groups do vary personality, and find that
composition affects performance \cite{duan2025power}, but they use small teams with no social
network and nothing to polarize about. The gap is explicit in the closest prior work:
Cau et al.\ \cite{cau2025selective} close their study of LLM opinion dynamics by naming
the absence of agent personalities and of network structure as its two principal
limitations. Consequently a question that sits directly between
them has not been answerable: \emph{does the personality composition of a society shape
polarization and collective intelligence at the same time, and do the two pull in opposite
directions?}

The question matters practically. If the compositions that keep a society talking across its
divisions are also the compositions that make it collectively stupid, then platform designers
and institution builders face a real dilemma, and interventions that reduce polarization
carry a hidden cost. If instead the two go together, no such trade-off needs managing. Nobody
has been well placed to say, because the two outcome families are rarely measured in the
same system.

We answer it with \emph{TraitMix}, a framework in which the Big Five composition of a
simulated online society --- both the average level of each trait and the diversity of traits
across agents --- is a controlled experimental variable, and in which polarization and
collective intelligence are measured in the same runs. Societies of one hundred agents,
carrying validated Big Five profiles, discuss contested political topics on a platform with
an algorithmic feed and a dynamic follow network, while being privately probed for their
opinions and asked to solve estimation and hidden-profile tasks with verifiable answers. We
report 991 such simulations.

\begin{figure}[t]\centering
\includegraphics[width=\linewidth]{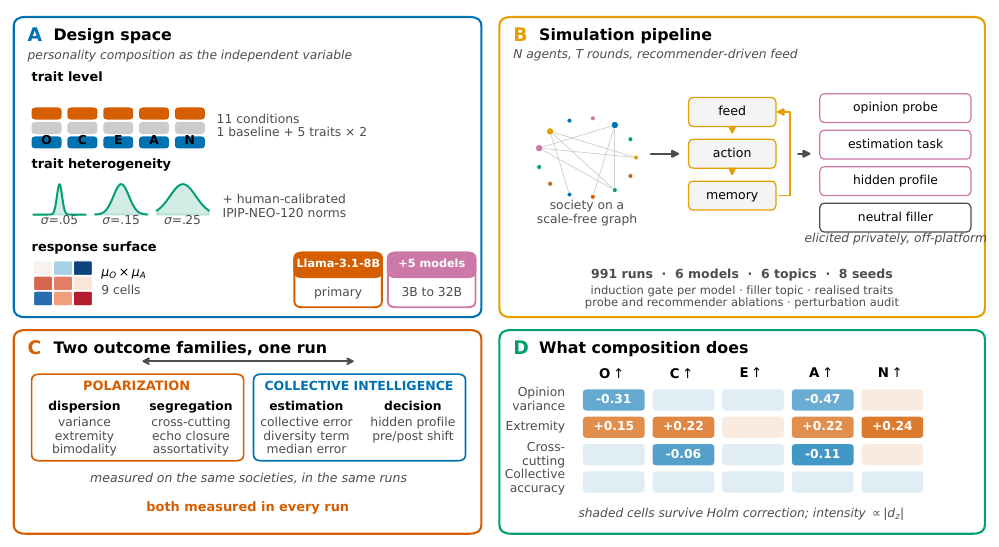}
\caption{Overview of the study.
\textbf{(A)} The design space. Personality composition is the controlled independent
variable: trait levels are varied one trait at a time, trait heterogeneity is varied with
levels held constant, and the two most consequential traits are crossed in a response
surface. A human-calibrated condition is built from published population norms.
\textbf{(B)} The simulation pipeline. Agents carrying sampled trait vectors populate a
scale-free follower network and interact for $T$ rounds through a recommender-driven feed,
while opinions and verifiable tasks are elicited privately, off the platform. The controls
listed beneath are reported in Section~\ref{sec:results}, including the measures that failed
them.
\textbf{(C)} The measurement architecture. Polarization is measured as dispersion and as
segregation, and collective performance through estimation and decision tasks, in the same
societies and the same runs.
\textbf{(D)} Which trait conditions moved which outcomes in the primary model. Shaded cells
survive Holm--Bonferroni correction; intensity is proportional to the standardised effect
size, and colour to its direction.}
\label{fig:overview}
\end{figure}

Figure~\ref{fig:overview} summarises the study. Three findings emerge. First, polarization is not one thing: the \emph{dispersion} of opinion
and its \emph{segregation} into camps respond differently to composition, and sometimes in
opposite directions, so that a homogeneous society turns out to be not a moderate one but a
consensual echo chamber. Second, the effect of one trait depends on another: Agreeableness
determines the sign of Openness, so that raising Openness increases polarization in
disagreeable societies and decreases it in agreeable ones --- an interaction that replicates
in a second model family and offers a concrete explanation for why single-trait findings in
this literature conflict. Third, and contrary to the trade-off we set out to measure,
polarization and collective intelligence are \emph{aligned}: societies that argue across their
divisions and retain diverse views are more accurate, not less.

Because a recent critical literature argues that validation is the central unsolved problem in
generative social simulation \cite{larooij2026validation, zhou2025pimmur}, we also treat the
controls as part of the contribution rather than as hygiene. A neutral filler topic, carried
in every society, was what allowed us to determine that one measure's failure to replicate
across model families was a response-style artifact rather than a substantive disagreement.

Our contributions are as follows.

\begin{enumerate}
\item \textbf{A framework in which personality composition is an experimental variable.}
TraitMix parameterises a society by the distribution of Big Five vectors across its agents,
separating trait level from trait heterogeneity, and instruments polarization and collective
intelligence in the same simulation runs. To our knowledge this is the first study to measure
both outcome families as a function of composition in a single system.

\item \textbf{Evidence that dispersion and segregation are distinct.} Across trait-level and
heterogeneity manipulations, opinion variance and extremity move independently of
cross-cutting interaction and echo-chamber closure, and personality diversity pushes them in
opposite directions. Reporting a single polarization statistic can therefore support opposite
conclusions about the same system.

\item \textbf{A replicated crossover interaction.} Agreeableness moderates the effect of
Openness on polarization, with a strongly negative interaction term recovered independently
in models of two families, explaining why single-trait accounts disagree.

\item \textbf{The absence of a trade-off, and one association that survives scrutiny.} No
polarization measure predicts poorer collective performance. Cross-cutting interaction
predicts greater collective accuracy, and is the only one of four polarization measures whose
association persists after partialling on both components of the diversity-prediction
identity and on realised trait dispersion; the other three do not survive that adjustment and
we withdraw them. The surviving association rests on variation between compositions rather
than within them, and we report it accordingly.

\item \textbf{A set of cheap, reportable validity controls, and an honest account of what
they caught.} A pre-specified induction gate, a neutral filler-topic control, exact
reconstruction of realised trait distributions, three unplanned internal replications, a
sign-stability audit across seven design perturbations, and a full cross-model replication.
We report the measures that failed these checks alongside those that passed, and release all
code, configurations and run-level data.
\end{enumerate}

The remainder of the paper is organised as follows. Section~\ref{sec:related} situates the
study between the two literatures. Section~\ref{sec:method} formalises the framework and
defines the outcome batteries and controls. Section~\ref{sec:setup} describes the models,
materials, experimental conditions and statistical protocol. Section~\ref{sec:results}
reports the results, and Section~\ref{sec:discussion} discusses their interpretation,
limitations and implications.

\section{Related work}\label{sec:related}

This study sits between two literatures that rarely meet. One studies how opinions polarize
in simulated online societies; the other studies how the composition of a group affects its
collective competence. Each has recently begun using large language models as agents, and
each has left the other's outcome unmeasured.

\subsection{Generative agent societies}

The modern line begins with generative agents: populations of language-model characters with
memory, reflection and planning, which produced recognisably social behaviour in a sandbox
environment \cite{park2023generative}. Subsequent platforms scaled the idea to social-media
settings with recommender-driven feeds and dynamic follow networks, up to very large agent
populations \cite{yang2024oasis, piao2025agentsociety}, embedded agents in richer
environments \cite{gao2023s3}, provided modular cognitive components for building such
simulations \cite{vezhnevets2023concordia}, and constructed benchmarks for the social
competence of the underlying models \cite{zhou2024sotopia}. The area has grown quickly enough
to be surveyed \cite{mou2026survey}.

Our simulation environment is deliberately conventional within this family: an interest- and
popularity-weighted feed, bounded agent memory, follow and unfollow actions, and free-text
posting. The contribution is not the environment but what is held fixed and what is varied
inside it.

\subsection{Opinion dynamics with language-model agents}

Classical opinion dynamics established that simple interaction rules generate consensus,
clustering or fragmentation depending on how tolerant agents are of disagreement
\cite{deffuant2000mixing}. Recent work re-runs these questions with language-model agents,
finding that such agents reproduce recognisable opinion dynamics
\cite{chuang2024simulating}, that echo chambers and human-like polarization can emerge from
networked interaction \cite{wang2025decoding, piao2025polarization}, and that alternative
feed algorithms change the character of the resulting discourse
\cite{tornberg2023simulating}.

Most directly relevant to the present study, Cau et al.\ \cite{cau2025selective} simulate
multi-round debates among LLM agents and find convergence toward agreement driven not by
sycophancy but by an asymmetric acceptance--rejection bias, together with a measurable role
for fallacious argument. Their agents, however, are deliberately uniform, and their
interactions are mean-field. The authors identify both as the principal limitations of their
framework, noting that agents ``lack distinct personalities or cognitive diversity'' and that
incorporating network structure such as clustering, assortativity and echo chambers would
improve realism. The present study takes up exactly these two extensions: personality
composition becomes the experimental variable, and interaction occurs over a dynamic network
whose assortativity and echo-chamber closure we measure. Relatedly, LLM populations have been
shown to develop shared conventions and collective bias through interaction alone
\cite{ashery2025emergent}, and algorithmic curation is known to amplify fragmentation in
classical opinion models \cite{sirbu2019algorithmic}.

Personality otherwise enters this literature only marginally. Where Big Five traits appear,
they are typically assigned as background colour to individuate agents rather than manipulated
as an experimental factor, and the societies studied are not compared across compositions
\cite{wang2025decoding}. Consequently the question of whether \emph{who the population is}
changes \emph{how it polarizes} has not been answered systematically. Nor, to our knowledge,
has any study in this line measured collective task performance alongside polarization.

\subsection{Personality in language models}

Whether a language model can be given a personality, and whether that personality can be
measured, has been examined directly. A psychometric framework for administering and shaping
Big Five traits in language models established that targeted trait levels can be induced and
recovered with substantial validity \cite{serapiogarcia2025psychometric}, and training on
human-grounded dialogue improves induction relative to prompting alone \cite{li2025big5chat}.
The instruments used are public-domain inventories from the International Personality Item
Pool \cite{goldberg1999ipip}, in particular the IPIP-NEO-120
\cite{johnson2014ipipneo120, maples2014ipip120}, whose structure has been examined in a very
large sample \cite{kajonius2019structure}.

This literature also supplies the main caution. Personality measurements in language models
are unstable across paraphrase, item order and conversation history \cite{tosato2026persist}, which
means an induction that validates on a questionnaire may not persist through extended
interaction. We take this seriously in Section~\ref{sec:limitations}: our own induction gate
is a questionnaire measure, and we do not claim more for it than that.

\subsection{Collective intelligence and group composition}

The collective-intelligence literature has long held that group accuracy depends on
independence and diversity of judgement rather than on individual expertise alone
\cite{page2008difference}, and that social influence can erode this by narrowing diversity
without improving accuracy \cite{lorenz2011social}. That conclusion is contested: influence
in decentralised networks can improve group estimates \cite{becker2017network}, and
deliberation within small independent groups can outperform aggregating a much larger crowd
\cite{navajas2018aggregated}. A separate classical result shows that discussion frequently
fails to pool information held by only one member, so groups choose the option favoured by
what everyone already knew \cite{stasser1985pooling, stasser1992discovery}.

Language models have entered this literature both as forecasters, where ensembles of models
rival human crowds \cite{schoenegger2024wisdom}, and as a subject of analysis, with recent
work considering how they might reshape collective intelligence more broadly
\cite{burton2024collective}. Personality composition has been studied in this setting, with
Big Five profiles assigned to agents in multi-agent teams and their effect on task
performance examined \cite{duan2025power}. These studies, however, use small teams without
a social network, and do not measure polarization: there is no opinion to polarize.

\subsection{Validity in generative social simulation}

A recent critical literature argues that this field's central unsolved problem is validation:
that simulations are frequently assessed on the believability of their transcripts rather
than against behavioural criteria \cite{larooij2026validation}, and that reported collective
phenomena can vanish or reverse when basic design principles are enforced \cite{zhou2025pimmur}.

We regard these criticisms as correct and have designed accordingly. The controls reported
in Section~\ref{sec:results-validation} --- a pre-specified induction gate, a neutral filler
topic to detect response-style artifacts, exact reconstruction of realised trait
distributions, unplanned internal replications, a sign-stability audit across design
perturbations, and full replication in a second model family --- are our response. As
Section~\ref{sec:results-crossmodel} shows, one of them changed what we were able to
conclude.

\begin{table}[t]\centering
\scriptsize\setlength{\tabcolsep}{2pt}
\caption{Scope conditions of this study and of related work on agent societies.}
\label{tab:related}
\begin{tabular}{@{}p{2.55cm}cccccc@{}}
\toprule
Study & Personality & Network & Polariz. & Coll.\ int. & Both & Cross-model \\
\midrule
\cite{park2023generative}      & --          & sandbox & --  & --  & -- & -- \\
\cite{gao2023s3}               & --          & yes     & yes & --  & -- & -- \\
\cite{tornberg2023simulating} & --          & yes     & yes & --  & -- & -- \\
\cite{cau2025selective}       & --          & mean-field & yes & -- & -- & partial \\
\cite{chuang2024simulating}   & --          & yes     & yes & --  & -- & partial \\
\cite{yang2024oasis}          & --          & yes     & yes & --  & -- & -- \\
\cite{piao2025polarization}   & --          & yes     & yes & --  & -- & -- \\
\cite{wang2025decoding}       & assigned    & yes     & yes & --  & -- & -- \\
\cite{schoenegger2024wisdom}  & --          & --      & --  & yes & -- & yes \\
\cite{piao2025agentsociety}    & --          & yes     & yes & --  & -- & -- \\
\cite{serapiogarcia2025psychometric} & manipulated & -- & --  & --  & -- & yes \\
\cite{duan2025power}          & manipulated & --      & --  & yes & -- & partial \\
\midrule
This work & manipulated & yes & yes & yes & yes & interaction only \\
\bottomrule
\end{tabular}
\par\smallskip\footnotesize
Studies are characterised only on the dimensions relevant to the question asked here, which
is not a summary of their overall contribution: several are landmark papers whose purpose was
different. ``Personality'' asks whether Big Five traits are an experimental factor rather
than background detail; ``assigned'' means profiles are given to agents but not varied across
conditions, ``manipulated'' means composition is itself a condition. ``Both together'' asks
whether polarization and collective intelligence are measured in the same system, which is
the gap this study addresses. ``partial'' under Cross-model means more than one model appears
somewhere in the study but the main result is not independently replicated in a second family.
Beyond the columns shown, this study separates trait \emph{level} from trait
\emph{heterogeneity}, distinguishes opinion dispersion from segregation, and reports a
pre-specified induction gate, a response-style control, a robustness audit across seven
design perturbations and a full cross-model replication. A recent review of this literature
concludes that validation is generally under-addressed \cite{larooij2026validation}, which is
the shortfall those controls are intended to answer.
\end{table}

\subsection{What is missing}

Table~\ref{tab:related} summarises the position. Across these literatures, personality
composition has been treated either as decoration or as a variable affecting task performance
in networkless teams; polarization and collective intelligence have been measured in separate
studies, on separate systems, by separate communities. We are not aware of a study that measures both outcome families in the same societies while
also manipulating personality composition, although the table records scope rather than
quality and the criteria were chosen for the question asked here.

This matters for more than completeness. If the two outcomes are measured separately, the
relationship between them is not merely unknown but unobservable, and the question that
motivates platform interventions --- whether reducing polarization costs collective
competence --- cannot be asked. Two further gaps follow from the same design choice. Because
prior work assigns personality rather than varying it, the possibility that one trait
determines the sign of another's effect has not been testable; and because most studies
report a single polarization statistic, the possibility that dispersion and segregation move
in opposite directions has not been visible.

Making these observable requires manipulating composition as an experimental factor,
measuring both outcome families in the same runs, and separating the two faces of
polarization, which is what we do.

\section{The TraitMix design}\label{sec:method}

We present \emph{TraitMix}, an experimental design in which the Big Five personality
composition of a simulated online society is a controlled independent variable, and in
which polarization and collective intelligence are measured as outcomes of the same
simulation run. This section formalises the framework, describes the agent architecture
and the personality-induction procedure, and defines the two outcome batteries together
with the artifact controls that accompany them. Panels A to C of Figure~\ref{fig:overview}
summarise the design space, the pipeline and the measurement architecture.

\subsection{Problem formulation}\label{sec:formulation}

\begin{definition}[Agent society]
A society is a directed graph $G=(V,E)$ with $|V| = N$ agents. Each agent $i \in V$ is
characterised by a Big Five trait vector
$\boldsymbol{\theta}_i = (\theta_i^{O}, \theta_i^{C}, \theta_i^{E}, \theta_i^{A},
\theta_i^{N}) \in [0,1]^5$, giving its levels of Openness, Conscientiousness,
Extraversion, Agreeableness and Neuroticism; a surface persona $p_i$ (name, age,
occupation); a bounded memory $M_i$; and, for each discussion topic $\tau$, a latent
opinion $x_i^{\tau}(t) \in \{-3,\dots,+3\}$ at round $t$.
\end{definition}

\begin{definition}[Personality composition]
The composition of a society is the distribution from which trait vectors are drawn,
$\boldsymbol{\theta}_i \sim \mathcal{TN}_5(\boldsymbol{\mu}, \Sigma)$, a five-dimensional
normal distribution truncated to $[0,1]^5$. The mean vector $\boldsymbol{\mu}$ specifies
trait \emph{levels} and $\Sigma$ specifies trait \emph{heterogeneity}. Composition is the
independent variable throughout: an experimental condition is a choice of
$(\boldsymbol{\mu}, \Sigma)$, with every other aspect of the simulation held fixed.
\end{definition}

Separating $\boldsymbol{\mu}$ from $\Sigma$ is deliberate. Two societies with identical
average personalities can differ in how varied their members are, and prior work on group
composition gives reasons to expect level and spread to act differently. Because
truncation couples them --- a distribution centred at $\mu=0.8$ with $\sigma=0.15$ is
clipped on one side and is therefore less dispersed than one centred at $\mu=0.5$ --- we
reconstruct and report the \emph{realised} trait moments for every run and use them as a
covariate (Section~\ref{sec:controls}).

\subsection{Agent architecture and interaction}\label{sec:agents}

Agents inhabit a social-media environment modelled on contemporary platforms and on the
generative agent-based simulators that preceded this work. Each round $t = 1,\dots,T$
proceeds as follows.

\emph{Activation.} Each agent independently becomes active with probability $\rho$,
reflecting the fact that most users of a platform are inactive at any given moment.

\emph{Feed construction.} An active agent $i$ receives a ranked feed of at most $f$ items
drawn from posts it has not authored. The ranking score of post $q$ combines a recency-
and popularity-driven term and an interest term,
\begin{equation}
s(q, i) \;=\; \Big( w_{\mathrm{int}} \cdot \underbrace{\big(1 - \tfrac{|x_i^{\tau(q)} -
x_{a(q)}^{\tau(q)}|}{6}\big)}_{\text{opinion proximity}} \;+\; w_{\mathrm{hot}} \cdot
\underbrace{\big(1 + \beta \ell_q + \max(0, \kappa - (t - t_q))\big)}_{\text{popularity and
recency}} \Big) \cdot \eta_{iq},
\label{eq:feed}
\end{equation}
where $a(q)$ is the author of $q$, $\tau(q)$ its topic, $\ell_q$ its like count, $t_q$ its
round of creation, and $\eta_{iq}>1$ a boost applied when $i$ follows $a(q)$. The interest
term in Equation~\ref{eq:feed} makes the recommender mildly homophilous, as engagement-optimising feeds are in
practice; because this is itself a design choice that could manufacture echo chambers, the
feed size and weights are perturbed in the robustness audit.

\emph{Action.} The agent is shown its feed and its recent memory, and selects exactly one
action from $\{\textsc{post}, \textsc{reply}, \textsc{like}, \textsc{follow},
\textsc{unfollow}, \textsc{pass}\}$, expressed in free text. Follow and unfollow actions
modify $E$, so the network is dynamic. Memory retains the last $k$ own-actions together
with any private information the agent holds.

\emph{Private probes.} Opinions are never inferred from public posts. Every $P$ rounds
each agent is asked privately, outside the platform, for its current position on each
topic on a $-3$ to $+3$ scale. Collective-intelligence tasks are likewise elicited
privately. This separation matters: it prevents an agent's stated position from being
confounded with its willingness to post, and it means polarization is measured over
beliefs rather than over expressed content.

\subsection{Personality induction and validation}\label{sec:induction}

Traits are induced by prompt. Each agent's system prompt states its persona and then
renders each trait as a graded natural-language description, using marker content from the
Big Five literature at one of nine levels from ``extremely low'' to ``extremely high''.
Agents are instructed to remain in character and never to refer to themselves as a
language model or to the simulation. Every prompt used in the simulation is reproduced
verbatim in Additional file~1.

Because induced personality cannot simply be assumed to have taken effect, induction is
validated before any experimental run, and separately for every model used. We administer the full 120-item IPIP-NEO-120
inventory, an instrument in the public domain, to every persona configuration used in the
main experiment, scoring items by their published keying and comparing measured trait
levels against targeted levels. This validation gate is reported in
Section~\ref{sec:results-validation}; the experimental programme was contingent on passing
it.

\emph{Rank agreement is not sufficient, and we report a magnitude criterion alongside it.}
Convergent validity in this literature is normally expressed as a correlation between
targeted and measured trait level, which asks whether configurations are ordered correctly.
It does not ask whether they are separated enough for the difference to matter. We encountered
a model that ordered nine configurations almost perfectly on Agreeableness ($\rho = 0.96$)
while separating the extremes by $0.71$ points on a five-point instrument, and whose
behaviour was correspondingly unaffected by composition. We therefore admit a model only if,
for each manipulated trait, the rank correlation reaches the pre-specified threshold
\emph{and} the difference in mean measured score between the highest and lowest targeted
level reaches one point of the instrument's five-point range.

We record that the magnitude criterion was added after observing that case, and was not
pre-specified. Three considerations bear on whether it is defensible. It is defined on a
quantity measured independently of any outcome, namely the questionnaire response. It is
applied identically to every model rather than selectively. And it agrees with an
independent behavioural quantity: the model it excludes on magnitude has a range of condition
means an order of magnitude below every admitted model. Both excluded models are reported in
full in Additional file~1 so that readers may judge the exclusions rather than accept them.

We further record induced personality \emph{behaviourally} during simulation. A trait
scorer is applied to each agent's public posts, and the mean absolute deviation between
expressed and targeted traits is logged for every run as a continuous manipulation check.
The distinction between questionnaire-based and behaviour-based validity is a live issue
in this literature, and we return to it in Section~\ref{sec:discussion}.

\subsection{Outcome batteries}\label{sec:outcomes}

Both outcome families are measured in every run, on the same societies, at the same time.
This is the central methodological commitment of the paper: prior work measures one or the
other, so the relationship between them has not been observable.

\subsubsection{Polarization}

For each contested topic $\tau$ with opinion vector $\mathbf{x}^\tau$ we compute:

\begin{itemize}
\item \textbf{Opinion variance} $\mathrm{Var}(\mathbf{x}^\tau)$ --- dispersion of belief.
\item \textbf{Extremity} $\frac{1}{N}\sum_i |x_i^\tau|$ --- average distance from neutrality.
\item \textbf{Bimodality}, via Sarle's coefficient and an Ashman-type separation statistic,
      capturing whether opinion has split into two camps rather than merely spread out.
\item \textbf{Opinion assortativity}, the numeric assortativity coefficient of
      $\mathbf{x}^\tau$ over $G$ --- whether agents are connected to the like-minded.
\item \textbf{Echo-chamber closure}, the negated Krackhardt E--I index over ties between
      agents of opposing opinion sign,
      $-\,(e_{\mathrm{ext}} - e_{\mathrm{int}})/(e_{\mathrm{ext}} + e_{\mathrm{int}})$,
      which is high when ties are predominantly within-camp.
\item \textbf{Cross-cutting interaction rate}, the fraction of replies exchanged between
      agents holding opposing opinion signs --- the behavioural, rather than structural,
      measure of whether the divide is being crossed.
\end{itemize}

We treat these as measuring two conceptually distinct things. Variance, extremity and
bimodality describe the \emph{dispersion} of opinion; assortativity, closure and
cross-cutting describe its \emph{segregation}. Our results show these can move in opposite
directions, so collapsing them into a single notion of ``polarization'' would obscure the
finding. We note one dependency: when opinion variance collapses, almost every tie is
within-camp by construction, so closure is partly mechanical at very low variance and is
never interpreted without variance alongside it.

\subsubsection{Collective intelligence}

Both task families are introduced into the conversation rather than posed only in private.
At each item's scheduled round an announcement post enters the feed, carrying likes so that
it competes in the recommender ranking, and agents may reply to it or discuss it in their own
posts. Exposure is therefore a measured quantity: across baseline runs, eight announcements
per run attract a mean of 34.7 direct replies and 13.0 further mentions, and 38 of the 100
agents engage with them. The private estimates elicited before and after are separated by a
window in which the item was available for discussion, and the finding that those estimates
barely move is a result about what the discussion achieved, not an artifact of the items
never having been seen.

Collective intelligence is measured on tasks with verifiable answers, embedded in the same
runs and elicited privately before and after a discussion window.

\emph{Numeric estimation.} Agents estimate real-world quantities whose true values are
published by the World Bank. For each item we compute, in $\log_{10}$ space, the error of
the collective estimate (the mean of individual estimates), the mean individual error, and
the diversity of estimates, which stand in the exact relation given by the
diversity-prediction decomposition: collective error equals mean individual error minus
prediction diversity. We report the collective error \emph{level} as the primary measure,
$z$-scored across items and sign-flipped so that larger values denote greater accuracy, and
also report the median relative error and the diversity term.

\emph{Hidden-profile decision.} Agents choose between two candidates. All agents share the
same eight facts, which favour the objectively weaker candidate; the facts establishing the
better candidate are unshared, each agent holding only a small random subset. No individual
can identify the correct choice alone, and only discussion that surfaces unshared
information can recover it. The task is announced on the platform and each agent's private
facts are placed in its memory, so that unshared information can enter the conversation.
We report the post-discussion correct rate.

We also computed the pre-to-post \emph{change} in estimation accuracy. As reported in
Section~\ref{sec:results-ci}, this measure proved unable to discriminate between
conditions, for a substantive reason we consider a finding in its own right rather than a
defect: private numeric estimates barely move across the discussion window. We report it
as a null.

\subsection{Artifact controls}\label{sec:controls}

Two alternative explanations are addressed by design rather than by argument.

\emph{Response-style priming.} Conditions with extreme trait levels contain stronger
language in their prompts, which might elicit more extreme answers on \emph{any} rating
scale, mimicking polarization. Every society therefore also holds a neutral, non-political
filler topic. If trait-extreme conditions merely answer everything more extremely, the
filler topic will move with the contested topics; if the effect is genuinely about
contested opinion, it will not.

\emph{Truncation of the trait distribution.} Because trait sampling is deterministic given
a condition and a random seed, the realised trait matrix of every completed run can be
reconstructed exactly, without re-simulation. We do so for all runs, report realised trait
means and standard deviations, and enter realised dispersion as a covariate when
interpreting effects of trait level.

\section{Experimental setup}\label{sec:setup}

\subsection{Models and serving}\label{sec:models}

The primary model is \textbf{Llama-3.1-8B-Instruct}. Five further models are used to test
whether the findings are specific to it: \textbf{Qwen2.5-3B}, \textbf{Qwen2.5-7B},
\textbf{Qwen2.5-14B} and \textbf{Qwen2.5-32B-Instruct}, the last three in AWQ quantisation,
and \textbf{Llama-3.2-3B-Instruct}. All three are served locally
through vLLM 0.8.5.post1 (PyTorch 2.6.0, CUDA 12.4) on a single NVIDIA RTX 4090 with 24\,GB
of memory, with a 4096-token context and 90\% GPU memory utilisation. Generation temperature
is 0.7 for agent actions and 0.3 for private probes, where determinism is preferable;
temperature is perturbed in the robustness audit.

\emph{How the models were chosen.} The selection is bounded by a deliberate constraint: the
entire study runs on one consumer graphics card, so that it can be reproduced by others at
modest cost rather than only by groups with cluster access. Within that limit we required
models that are open-weight, instruction-tuned, in current use, and servable reliably enough
that data integrity could be guaranteed. Six were run, spanning 3 to 32 billion parameters
and two families, and they vary the two properties that matter for the question at hand:
family, which tests whether an effect is a quirk of one developer's training pipeline, and
parameter count, which tests whether it depends on scale. Four of the six sizes fall within a
single family, giving within-family size contrasts at constant quantisation, so a difference
between them cannot be attributed to family or to numerical format.

We emphasise what this design is not. It is not a benchmark, and we make no claim to have
surveyed the space of language models. The design is also unbalanced: the two families are
not represented at matching sizes throughout, so family and scale are only partially
separable. Additional models were attempted and not included, either because they could not
be served within the memory available or, in one case, because the inference server returned
errors during the run that we could not rule out as having affected the generated data; we
discarded that data rather than report it. Two further models were run to completion but
excluded by induction validation, and are reported in Additional file~1. The claim we make
from the remainder is correspondingly narrow, and stated in
Section~\ref{sec:results-crossmodel}. Testing the framework on any further model requires one
command in the released code, and we would welcome that.

Serving locally, rather than through a commercial API, is a deliberate choice: it fixes the
model version for the lifetime of the study, it makes the entire pipeline reproducible by
others at modest cost, and it avoids silent provider-side model updates that would compromise
a controlled experiment. The exact environment, including package versions, is released with
the code.

\subsection{Society configuration}\label{sec:society}

\begin{figure}[p]\centering
\includegraphics[width=\linewidth]{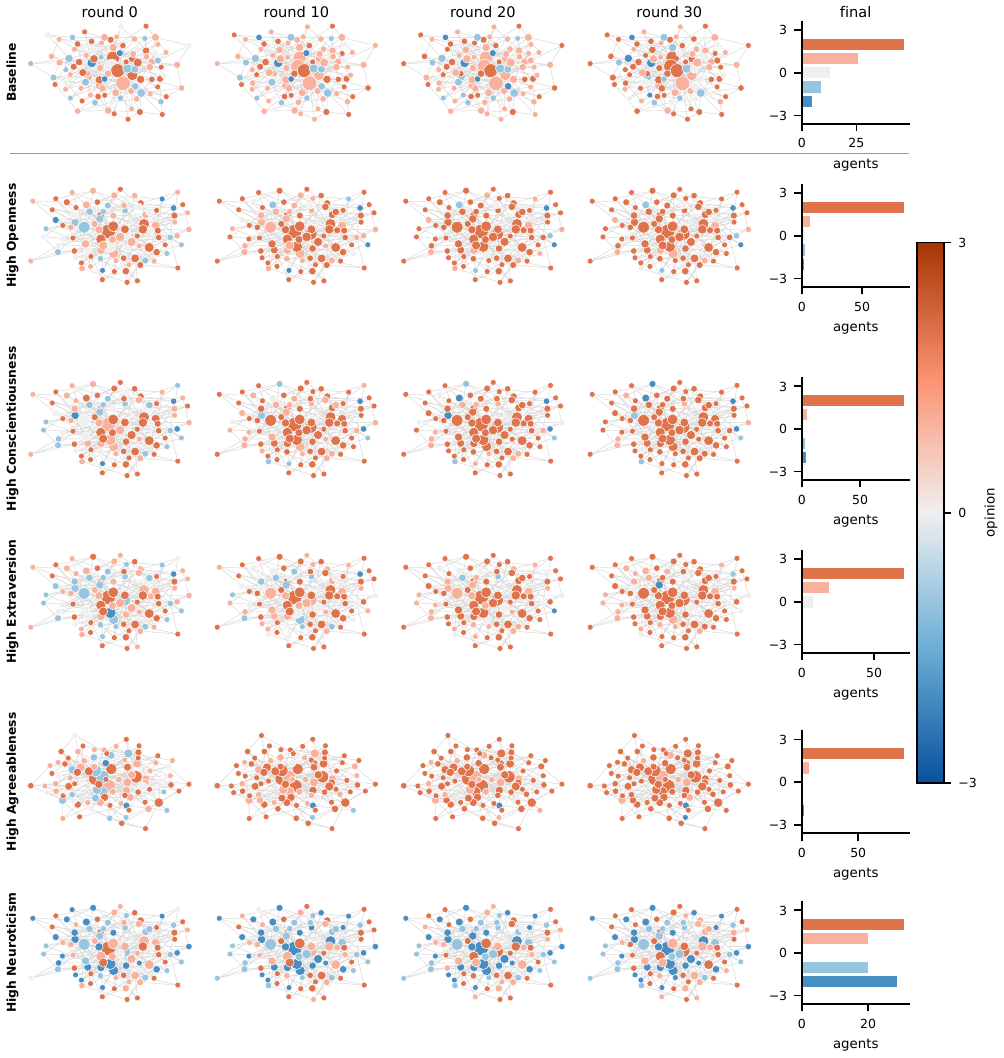}
\caption{The simulated society over thirty rounds, for the baseline and each raised trait.
Each panel shows the same hundred agents on their follower network, an agent's colour giving
its privately held opinion on gun control at that round and its size scaling with degree. The
right-hand column gives the final distribution of opinion. One representative run is shown
per condition: final opinion variance in the runs displayed is $1.42$, $0.70$, $0.88$,
$0.59$, $0.37$ and $2.83$ from top to bottom, against condition means over eight seeds of
$1.34$, $0.73$, $0.77$, $0.83$, $0.30$ and $2.71$. Convergence is the ordinary outcome: four
of the five raised traits produce societies that settle onto one side, most completely under
high Agreeableness, which ends at a $98$:$2$ split. High Neuroticism is the exception,
sustaining two camps of near-equal size ($51$:$49$). Ties shown are the network as
initialised; agents also follow and unfollow during a run.}
\label{fig:society}
\end{figure}

Unless stated otherwise, societies comprise $N = 100$ agents over $T = 30$ rounds with
activation probability $\rho = 0.4$, feed size $f = 10$, memory $k = 10$, and private
opinion probes every $P = 5$ rounds. The initial network is a Barabási--Albert graph with
$m = 3$, giving the heavy-tailed degree distribution characteristic of social platforms;
alternative topologies are examined in the audit. Figure~\ref{fig:society} shows a
society of this kind evolving over the course of a run, for the baseline and for each
trait raised in turn. Feed weights are
$w_{\mathrm{int}} = 1.0$ and $w_{\mathrm{hot}} = 0.5$.

The choice of $N = 100$ warrants a word. Statistical power in a factorial design of this
kind resides in the number of conditions and seeds, not in the number of agents, because
each society contributes a single observation per network-level metric. One hundred agents
is sufficient for stable network statistics, is at or above the operating scale of
comparable controlled studies, and keeps a single run to roughly three minutes, which is
what makes a programme of this size feasible on one consumer GPU. A scale ablation at $N = 200$
tests whether conclusions are scale-dependent.

\subsection{Materials}\label{sec:materials}

\emph{Personality inventory.} The IPIP-NEO-120, comprising 120 items across 30 facets with
24 items per Big Five domain and both positively and negatively keyed items. The items are
in the public domain. We derived our machine-readable item file directly from the published
keying and verified that it contains exactly 120 items, 24 per domain, 30 facets, and no
duplicates before use.

\emph{Discussion topics.} Two contested policy statements (national gun-control law;
immigration) plus one neutral filler statement used as the response-style control described
in Section~\ref{sec:controls}.

\emph{Estimation items.} Candidate items were real-world quantities with World Bank
ground truth. Items were \emph{screened before the experiment} against two pre-specified
criteria: the model's solo median relative error must be at least 15\%, so that there is
headroom for a crowd to be right or wrong, and the standard deviation of $\log_{10}$
estimates must be at least 0.02, so that agents genuinely disagree and the diversity term
is non-degenerate. Of fourteen candidates, six passed, and the full screening record including rejected items
is given in Additional file~1. The screen is not a formality: items
such as the population of Kenya were rejected because every agent returned a byte-identical
answer with 3\% error, leaving no aggregation for the crowd to perform. The full screening
record, including rejected items and their statistics, is released with the data.

\emph{Hidden-profile task.} Constructed as described in Section~\ref{sec:outcomes}, with
eight shared facts favouring the weaker candidate, six unshared facts favouring the better
candidate, three unshared facts disfavouring the weaker one, and two unshared facts per
agent.

\emph{Normative personality statistics.} The human-calibrated condition uses domain means
and standard deviations from a published open-access study of the IPIP-NEO-120 in a sample
of 320{,}128 respondents, rescaled from the instrument's 4--20 domain range to $[0,1]$. We
follow the original authors in noting that this sample, though large, consists of
self-selected volunteers and is not nationally representative; the condition is therefore
described as human-\emph{calibrated} rather than human-representative.

\subsection{Experimental conditions}\label{sec:conditions}

The programme comprises 991 simulation runs. Table~\ref{tab:runaccounting} maps every
condition prefix to its experiment, model and topic set, so that the total can be
reconstructed from the released data.

\begin{table}[t]\centering
\caption{Run accounting. Condition prefixes map to experiment, model and topic set; the run counts sum to the total reported in the article.}
\label{tab:runaccounting}
\scriptsize
\setlength{\tabcolsep}{4pt}
\begin{tabular}{@{}lp{3.2cm}p{3.0cm}p{1.8cm}rr@{}}
\toprule
Prefix & Experiment & Model & Topics & Conditions & Runs \\
\midrule
\texttt{e1t} & E1 six-topic replication & Llama-3.1-8B & 6 topics & 11 & 88 \\
\texttt{e1} & E1 trait levels & Llama-3.1-8B & 2 topics & 11 & 88 \\
\texttt{e2} & E2 heterogeneity & Llama-3.1-8B & 2 topics & 4 & 32 \\
\texttt{e2t} & E2 six-topic replication & Llama-3.1-8B & 6 topics & 4 & 32 \\
\texttt{e3} & E3 response surface & Llama-3.1-8B & 2 topics & 9 & 72 \\
\texttt{e5} & E5 robustness audit & Llama-3.1-8B & 2 topics & 24 & 72 \\
\texttt{abpr} & R1 probe-anchor ablation & Llama-3.1-8B & 2 topics & 10 & 80 \\
\texttt{abex} & R2 expressed-stance variant & Llama-3.1-8B & 2 topics & 3 & 24 \\
\texttt{abwi} & R2 recommender ablation & Llama-3.1-8B & 2 topics & 10 & 80 \\
\texttt{scale} & Scale ablation & Llama-3.1-8B & 2 topics & 1 & 3 \\
\texttt{e3llama3b} & E3 replication & Llama-3.2-3B (excluded) & 2 topics & 9 & 45 \\
\texttt{e1q} & E1 replication & Qwen2.5-14B & 2 topics & 11 & 55 \\
\texttt{e2q} & E2 replication & Qwen2.5-14B & 2 topics & 4 & 20 \\
\texttt{e3q} & E3 replication & Qwen2.5-14B & 2 topics & 9 & 45 \\
\texttt{e3l} & E3 replication & Qwen2.5-32B & 2 topics & 9 & 45 \\
\texttt{e3qwen3b} & E3 replication & Qwen2.5-3B (excluded) & 2 topics & 9 & 45 \\
\texttt{e1qwen7} & E1 replication & Qwen2.5-7B & 2 topics & 11 & 88 \\
\texttt{e2qwen7} & E2 replication & Qwen2.5-7B & 2 topics & 4 & 32 \\
\texttt{e3qwen7} & E3 replication & Qwen2.5-7B & 2 topics & 9 & 45 \\
\midrule
& & & & \textbf{162} & \textbf{991} \\
\bottomrule
\end{tabular}
\end{table}

\textbf{E0 --- Induction validation.} The IPIP-NEO-120 administered to all persona
configurations used in E1, three replicates each, before any experimental run.

\textbf{E1 --- Trait levels.} Eleven conditions: a baseline with all trait means at 0.5,
plus each of the five traits set to 0.8 and to 0.2 in turn, with all others held at 0.5 and
$\sigma = 0.15$ throughout. Eight seeds.

\textbf{E2 --- Trait heterogeneity.} Trait means fixed at 0.5, with $\sigma \in \{0.05,
0.15, 0.25\}$, plus the human-calibrated condition. Eight seeds. This isolates spread from
level.

\textbf{E3 --- Response surface.} A $3\times3$ grid crossing Openness and Agreeableness
means at $\{0.2, 0.5, 0.8\}$, the two traits identified as most consequential in E1. Eight
seeds.

\textbf{E5 --- Robustness audit.} (The numbering follows the pre-specified design, in
which a fourth experiment comparing prompt-based with fine-tuning-based induction was
planned. It was not run, because the induction gate was passed by prompting alone and the
arm was insurance against its failure. The label is retained so that experiment names match
the released configuration files and results.) Two anchor conditions, each crossed with seven design perturbations --- Watts--Strogatz and Erdős--Rényi topologies, activation probability
halved and increased by half, memory doubled, feed size increased by half, and temperature
raised to 1.0 --- plus a model swap. Three seeds. The audit is pre-specified in the sense
that the perturbation set was fixed before results were examined; its role is sign
stability rather than significance testing, and it is analysed accordingly
(Section~\ref{sec:stats}).

\textbf{Cross-model replication.} E1, E2 and E3 repeated in full on Qwen2.5-14B with
byte-identical compositions, five seeds.

\textbf{E1/E2 six-topic replication.} The trait-level and heterogeneity conditions
repeated with six contested topics instead of two, retaining the original pair, at eight
seeds. This tests whether topic-averaged effects are properties of composition or of the
topic set.

\textbf{Circularity ablations.} Ten conditions repeated with the probe anchors removed,
and the same ten with the recommender's opinion-proximity term removed, at eight seeds each;
and three conditions with the proximity term computed from an agent's most recently posted
stance rather than its latent opinion, which is the recommender a platform could actually
build. These test whether the measurement instrument or the feed generates the effects they
are used to measure.

\textbf{Scale ablation.} The baseline condition at $N = 200$, three seeds.

Three conditions --- the E1 baseline, the mid-heterogeneity condition of E2, and the centre
cell of the E3 grid --- have identical compositions but were executed as separate
conditions with independent seeds. They constitute an unplanned internal replication and we
report their agreement as a reliability check.

\subsection{Statistical protocol}\label{sec:stats}

The protocol was specified before the main runs and is reported in full because two of its
provisions materially affect what can be claimed.

\emph{Units and inference.} Each run contributes one observation per metric. Seeds are
matched across conditions, so contrasts against the baseline are paired. Primary inference
uses paired $t$-tests on run-level values, supported by linear mixed-effects models on
run\,$\times$\,topic and run\,$\times$\,item observations with seed as a random intercept.

\emph{A note on the nonparametric alternative.} The Wilcoxon signed-rank test with $n$
paired observations cannot return a two-sided $p$ below $2/2^{n}$: 0.0625 at five seeds and
0.0078 at eight. With multiplicity correction across ten or more contrasts it is therefore
\emph{mathematically incapable} of reaching significance at these sample sizes, and a
verdict of ``nothing is significant'' from it would be an artifact of the test rather than
evidence of absence. We report it as confirmatory only and state the floor explicitly.

\emph{Multiplicity.} Holm--Bonferroni correction is applied within each experiment family
$\times$ metric. Families are corrected separately because they answer separate questions.

\emph{Effect sizes.} Cohen's $d_z$ for paired data, Hedges' $g$, and Cliff's $\delta$ are
reported for every contrast regardless of significance, with bootstrap 95\% confidence
intervals on the mean difference (10{,}000 resamples). Contrasts with large effect sizes
that do not survive correction are reported as power-limited rather than null.

\emph{Seed extension.} A single pre-specified extension was permitted: contrasts falling
between $0.05$ and $0.15$ after correction at five seeds triggered extension to eight. When
this rule fired, all conditions in the affected families were extended, not only the
flagged ones, because paired tests require the baseline to be extended in step and
selective extension would have capped power silently.

\emph{Baselines under replication.} Conditions run on the replication model are contrasted
against the \emph{replication-model} baseline. Differencing a Qwen condition against a
Llama baseline would measure the model, not the composition.

\emph{Exclusion of the audit from significance testing.} At three seeds the audit's
$t$-tests produce uninterpretable effect sizes; the audit is analysed exclusively through
sign stability, defined as whether an anchor's difference from the baseline preserves its
direction when both are subjected to the same perturbation.

\emph{Standardisation is within model.} The collective-intelligence measures are composites
of per-item scores, and per-item scores are on incomparable scales, so each is standardised
before averaging. The reference population for that standardisation is the set of runs from
the same model, not the dataset as a whole. This matters for reproducibility rather than for
interpretation: standardising against the whole dataset makes a reported statistic depend on
which other models happen to be present, so the same runs yield different values as a dataset
grows, and a reader recomputing the number from the released data may not recover it. Within
one model the reference is fixed by the analysis itself, and we verified that the reported
associations are unchanged when models are added to or removed from the released results.

\subsection{Validity commitments and implementation}\label{sec:validity}

The design follows recent proposals for validity in LLM-based social simulation. Agent
profiles are heterogeneous by construction and validated rather than asserted; interaction
occurs through a real feed rather than by broadcast; agents carry bounded memory; action
prompts are open-ended and contain no instruction that presupposes an outcome; agents are
never told they are in an experiment; and a human-calibrated condition anchors the design
to observed population statistics.

All numbers reported in this paper are means over seeds with standard deviations, produced
by an analysis pipeline that reads the run-level results file and emits every table and
figure automatically. No value was transcribed by hand. Additional file~1 contains a results ledger:
one row for every quantity quoted in this article, giving the file, column and
aggregation that produced it, so that any number here can be recomputed from the
released run-level data. The simulation framework, the
configuration files defining every condition, the analysis and figure code, and the complete
run-level results are released so that the entire programme can be reproduced end to end.

\emph{Use of large language models.} Beyond their role as the object of study, large
language models were used as a writing and coding aid during the preparation of this work.
All study design, analysis and interpretation are the authors' own, and the authors take full
responsibility for the content of this manuscript. No language model is listed as an author.

The pipeline includes two integrity safeguards. Configurations that depend on normative
statistics will not execute until those statistics are supplied together with their source,
which prevents placeholder values from entering a run. The model client aborts a run when
the proportion of failed model calls in a batch exceeds a threshold, rather than substituting
null actions, since a run completed under a misconfigured server would otherwise yield a
complete but empty results record. A screening procedure for this failure mode is released
with the code, and the reported dataset was verified against it.

\section{Results}\label{sec:results}

\begin{figure}[t]\centering
\includegraphics[width=0.86\linewidth]{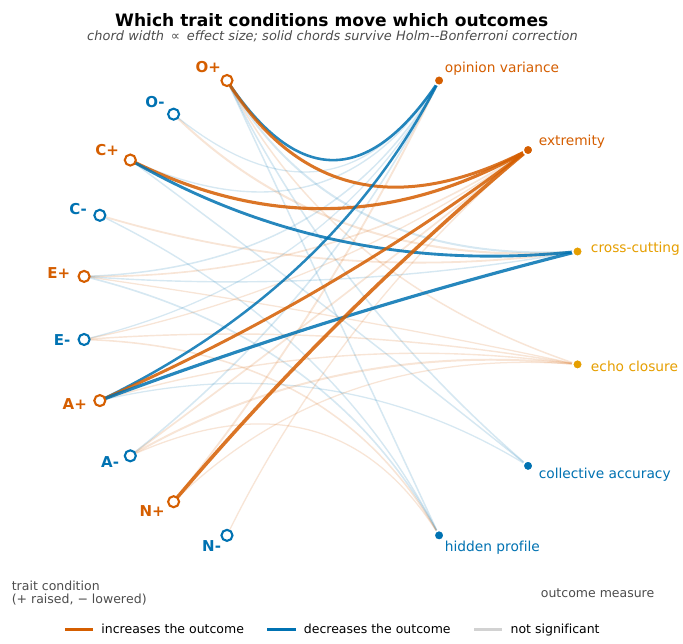}
\caption{Map of trait-level effects on both outcome families.
Each chord links a trait condition, on the left, to an outcome measure, on the right. Chord
width is proportional to the standardised effect size and colour to its direction, with solid
chords marking contrasts that survive Holm--Bonferroni correction within family and metric
and faint chords those that do not. The asymmetry between the upper and lower halves is the
substance of Sections~\ref{sec:results-levels} to \ref{sec:results-surface}: raised traits
move polarization measures strongly and in opposing directions, while effects on collective
intelligence are sparse and weak. Numerical values, confidence intervals and exact $p$ values
for every contrast shown here are given in Table~\ref{tab:e1} and Figure~\ref{fig:forest},
and for all contrasts in Additional file~1.}
\label{fig:effectmap}
\end{figure}

We report 991 simulation runs. Section~\ref{sec:results-validation} establishes that the
personality manipulation took effect and that the design is free of the two artifacts it was
built to detect. Sections~\ref{sec:results-levels} to \ref{sec:results-surface} give the
effects of trait level, trait heterogeneity, and their interaction.
Sections~\ref{sec:results-ci} and \ref{sec:results-align} turn to collective intelligence and
to the relationship between the two outcome families.
Sections~\ref{sec:results-robust} and \ref{sec:results-crossmodel} report what survives
design perturbation and a change of model family. Figure~\ref{fig:effectmap} maps which trait conditions moved which outcomes, and the
subsections that follow report each in detail. Unless stated otherwise, values are means
over eight seeds, differences are contrasts against the same-model baseline, and $p$ values
are Holm--Bonferroni corrected within experiment family and metric.

\subsection{Manipulation and artifact checks}

\emph{Two circularity checks.} Two features of the design could in principle generate the
results they are used to measure, and each was removed and the experiment repeated. The
private opinion probe stated the agent's previous answer and the average of the posts it had
seen, which is an influence operation inside the measurement instrument; the recommender
ranked posts partly by the distance between two agents' latent opinions, which is the same
variable the polarization measures are computed on. We re-ran ten conditions with each
feature removed, 160 runs in total, and compare the resulting condition effects against the
published ones.

Removing the opinion-proximity term from the recommender changes almost nothing: across four
measures and nine conditions the rank agreement with the published effects is $\rho = 0.975$
and 33 of 36 effects keep their sign. Echo-chamber closure and cross-cutting interaction, the
two measures most exposed to the objection, are among the best preserved. The segregation we
report is therefore not manufactured by the feed.

Removing the probe anchors is less inert, and we report the difference rather than claim the
ablation had no effect. Rank agreement is $\rho = 0.729$ with 29 of 36 signs preserved; the
effects the article leads on are intact (high Agreeableness on opinion variance $-0.465$
against $-0.456$ published; the homogeneous condition $-0.772$ against $-0.521$), the
Openness by Agreeableness crossover is preserved and larger, and the heterogeneity
dose-response is slightly stronger without the anchors ($r = +0.981$ against $+0.966$). What
changes is extremity, which rises when the anchors are removed, and three low-magnitude
conditions that flip sign. The anchors were therefore compressing opinions toward the stated
feed average, as an anchoring account would predict, but they were not generating the
findings. Full comparisons are given in Additional file~1.

\label{sec:results-validation}

\emph{Induction succeeded.} Across the persona configurations used in the main experiment,
targeted and measured trait levels correlated at $\rho = 0.93$ on average (Openness $0.89$,
Conscientiousness $0.92$, Extraversion $0.95$, Agreeableness $0.95$, Neuroticism $0.96$;
Table~\ref{tab:e0}). The gate set in advance was $\rho \geq 0.60$, so the experimental
programme proceeded. We return in Section~\ref{sec:discussion} to what a correlation this
high does and does not establish.

\emph{The effects are not response-style priming.} In the primary model, variance on the
neutral filler topic was unrelated to extremity on the contested topics ($r = -0.061$,
$p = 0.33$, $n = 258$). Priming would produce a positive association; the observed sign is
negative and the relationship is absent. The clearest single case is the high-Neuroticism
condition, which produced the highest contested-topic extremity in the study alongside among
the lowest filler-topic variance. Polarization here is specific to contested content.

\begin{figure}[t]\centering
\includegraphics[width=\linewidth]{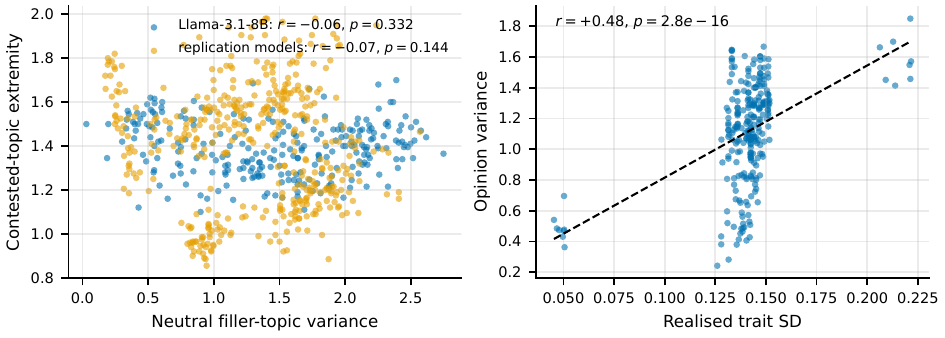}
\caption{Artifact controls. \textbf{Left:} variance on the neutral filler topic against
extremity on the contested topics, by model family. In the primary model the two are
unrelated, ruling out response-style priming; in the replication model they are strongly
associated, which Section~\ref{sec:results-crossmodel} shows is why one measure fails to
replicate there. \textbf{Right:} realised trait dispersion against opinion variance in the
primary model. Dispersion predicts variance, but accounts for roughly one twelfth of the
largest condition effect.}
\label{fig:controls}
\end{figure}

\emph{Truncation contributes, but does not explain the effects.} Realised trait dispersion
predicts opinion variance across runs ($r = 0.480$, $p < 10^{-15}$, $n = 258$;
Figure~\ref{fig:controls}, right, and Additional file~1), and the
relationship persists with condition held constant ($b = 6.67$, $p < 10^{-4}$, model
$R^2 = 0.857$) --- a within-condition natural experiment in which seeds that happened to draw
more varied personalities produced more varied opinions. The magnitude is nonetheless modest
relative to the effects of trait level. Between the baseline and the least polarized cell of
the response surface, realised dispersion differs by $0.011$, which the fitted coefficient
converts into an expected difference in opinion variance of roughly $0.07$ against an
observed difference of $0.84$: about one twelfth. Two conditions run directly against the
confound. High Neuroticism has below-baseline dispersion yet the highest opinion variance in
the study, and the cell combining high Openness with low Agreeableness has nearly the lowest
dispersion together with among the highest variance. We therefore interpret trait level as
acting in its own right, and report realised dispersion as a covariate throughout.

\emph{Three independent runs of the same composition agree.} The E1 baseline, the
mid-heterogeneity condition of E2, and the centre cell of the E3 grid are compositionally
identical but were executed as separate conditions with independent seeds. Their means
differ by at most $0.052$ on opinion variance, $0.039$ on extremity, $0.032$ on cross-cutting
rate and $0.007$ on hidden-profile accuracy, against condition differences spanning more than
an order of magnitude. This unplanned replication indicates that run-to-run reliability is
high relative to the effects being measured.

\begin{table}[t]\centering
\small
\caption{Induction validation (E0). Spearman correlation between targeted and measured trait level, IPIP-NEO-120 administered to every persona configuration used in the main experiment.}
\label{tab:e0}
\begin{tabular}{lcccccc}
\toprule
Induction & O & C & E & A & N & Mean $r$ \\
\midrule
prompt & 0.89 & 0.92 & 0.95 & 0.95 & 0.96 & \textbf{0.93} \\
\bottomrule
\end{tabular}
\end{table}

\subsection{Trait levels shape how societies disagree}\label{sec:results-levels}

\begin{figure}[t]\centering
\includegraphics[width=\linewidth]{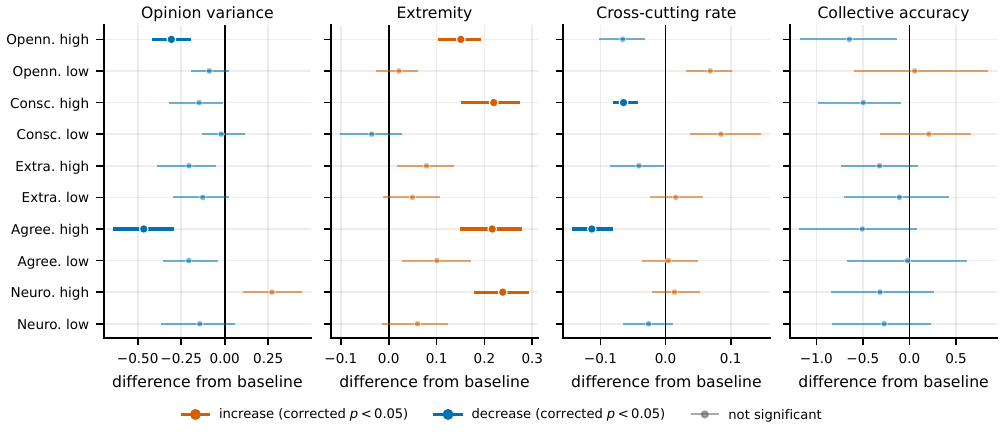}
\caption{Trait-level effects with bootstrap confidence intervals (E1). Each point is the
difference between a condition and the baseline, averaged over eight seeds, with a 95\%
bootstrap interval. Filled, heavier markers denote contrasts surviving Holm--Bonferroni
correction within family and metric. Raising Agreeableness lowers opinion variance and
cross-cutting interaction while raising extremity; raising Neuroticism raises both variance
and extremity. Collective accuracy shows no corrected-significant trait-level effect.}
\label{fig:forest}
\end{figure}

Table~\ref{tab:e1} and Figure~\ref{fig:forest} report the trait-level contrasts; ten
survive correction. Every contrast tested, significant or not, is listed in
Additional file~1.

\begin{figure}[t]\centering
\includegraphics[width=\linewidth]{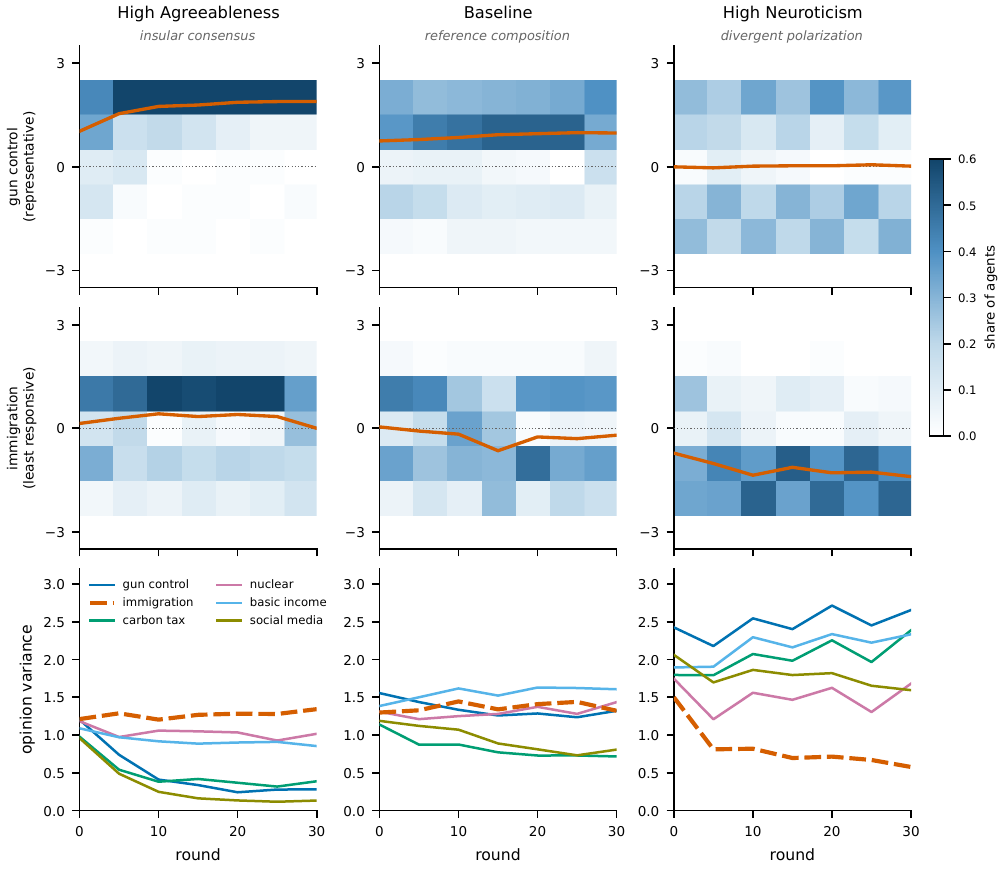}
\caption{Two regimes of opinion dynamics, and the one topic that does not follow them.
The top two rows show one representative six-topic society per condition, the same society
in both rows, with the density of agent opinions at each round and the population mean in
orange. The first row shows gun control, a topic on which composition has strong purchase;
the second shows immigration, the topic on which it has least. The bottom row gives the mean
within-run opinion variance for all six contested topics. Under high Agreeableness five
topics converge toward consensus while immigration remains dispersed; under high Neuroticism
five diverge while immigration alone converges. The two regimes described in the text are
therefore properties of composition that hold across most topics, with immigration the
consistent exception analysed in Section~\ref{sec:results-topic}.}
\label{fig:dynamics}
\end{figure}

\emph{The segregation measures are rates, so we report their denominators.} A
cross-cutting rate near zero can arise in two ways: because agents declined to reply across
a divide, or because a society had converged so far that few opposing pairs remained. These
are different claims and the ratio alone cannot separate them. Recovering the counts from
the post record shows the first rather than the second. Under high Agreeableness, agents
exchanged 598 replies, more than the 435 exchanged at baseline, and 4,069 ordered pairs of
agents held opposing-sign opinions, against 4,892 at baseline; 96\% of agents still held a
non-zero opinion. The opportunity to reply across the divide was present and was not taken:
25 of 598 replies crossed it, against 72 of 435 at baseline. The homogeneous condition is
the one where the pool of opposing pairs genuinely thins, to 14.5\% of possible pairs, and
we mark that rate as resting on a smaller base. Full counts for five conditions are given in
Additional file~1.

Raising \textbf{Agreeableness} produces the most consequential single change. Cross-cutting
interaction falls by $0.111$ (95\% CI $[-0.141, -0.078]$, $d_z = -2.37$, $p = 0.003$) and
opinion variance by $0.476$ ($d_z = -1.73$, $p = 0.016$), while extremity \emph{rises} by
$0.213$ ($d_z = 2.12$, $p = 0.004$). Agreeable societies do not become moderate. They
converge, and they converge on a more extreme position, while largely ceasing to argue
across the divide.

Raising \textbf{Conscientiousness} does something similar but weaker: cross-cutting falls by
$0.062$ ($d_z = -2.12$, $p = 0.005$) and extremity rises by $0.216$ ($d_z = 2.38$,
$p = 0.002$). Raising \textbf{Openness} lowers opinion variance by $0.318$ ($d_z = -1.83$,
$p = 0.013$) and raises extremity by $0.147$ ($d_z = 2.22$, $p = 0.003$).

Figure~\ref{fig:dynamics} shows these two regimes forming: under high
Agreeableness the opinion distribution contracts onto a single position, while under
high Neuroticism it splits into two camps that persist to the end of the run.

Raising \textbf{Neuroticism} produces the opposite pattern. It yields the largest extremity
increase in the family ($0.234$, $d_z = 2.75$, $p = 0.001$) together with the highest opinion
variance of any E1 condition. Where agreeable and open societies converge on extremes,
neurotic societies pull apart on them.

\emph{An asymmetry we previously reported does not survive the wider topic set.} In the
two-topic design, all ten corrected-significant trait-level contrasts came from conditions in
which a trait was raised, and none from conditions in which one was lowered. We reported that
asymmetry as a finding. It does not hold. Repeating the same conditions with six topics,
which raises the number of observations entering each contrast, 17 of 20 raised contrasts and
14 of 20 lowered contrasts survive correction. The asymmetry was a consequence of low power
in a design with two topics, not a property of composition, and we withdraw it.

This is the first of two places where treating polarization as a single quantity would have
obscured the result. Dispersion (variance, extremity) and segregation (cross-cutting, echo
closure) do not move together, and conditions that lower one can raise the other.

That claim needs more than an inspection of signs, because the measures are not independent
by construction: when opinion variance collapses, few pairs of agents hold opposing positions,
so cross-cutting replies become scarce and echo-chamber closure rises mechanically. The
measures are indeed correlated in these data ($r = +0.74$ between opinion variance and
cross-cutting, $r = -0.76$ between variance and echo closure). We therefore test whether
composition moves the segregation measures beyond what opinion variance alone implies, by
regressing each on opinion variance and asking what condition adds. The addition is substantial:
$R^2 = 0.60$ for cross-cutting ($F(45,211) = 7.10$, $p < 10^{-22}$), $0.63$ for echo-chamber
closure ($F = 7.96$, $p < 10^{-25}$) and $0.83$ for extremity ($F = 23.55$, $p < 10^{-59}$).
The measures are related but not redundant, and the dissociation is a property of the data
rather than of the choice of statistic.

\begin{table}[t]\centering
\scriptsize\setlength{\tabcolsep}{1.5pt}
\caption{Effects of trait level on polarization and collective intelligence (E1).}
\label{tab:e1}
\begin{tabular}{@{}lccccc@{}}
\toprule
Condition & Opinion var. & Extremity & Cross-cut & Echo closure & Coll. accuracy \\
\midrule
\textit{norm baseline} (baseline) & 1.263\,{\scriptsize$\pm$0.209} & 1.296\,{\scriptsize$\pm$0.077} & 0.165\,{\scriptsize$\pm$0.039} & 0.361\,{\scriptsize$\pm$0.163} & 0.106\,{\scriptsize$\pm$0.689} \\
\midrule
agreeableness high & \textbf{-0.465$^{*}$} & \textbf{+0.216$^{**}$} & \textbf{-0.113$^{**}$} & +0.146 & -0.508 \\
agreeableness low & \textbf{-0.207} & \textbf{+0.101} & +0.004 & \textbf{+0.209} & -0.023 \\
conscientiousness high & -0.148 & \textbf{+0.219$^{**}$} & \textbf{-0.064$^{**}$} & +0.063 & -0.496 \\
conscientiousness low & -0.020 & -0.036 & \textbf{+0.085} & -0.011 & +0.207 \\
extraversion high & -0.206 & \textbf{+0.079} & -0.041 & +0.115 & -0.322 \\
extraversion low & -0.127 & +0.049 & +0.016 & +0.141 & -0.109 \\
neuroticism high & \textbf{+0.271} & \textbf{+0.238$^{**}$} & +0.014 & +0.105 & -0.317 \\
neuroticism low & -0.144 & +0.060 & -0.026 & +0.070 & -0.272 \\
openness high & \textbf{-0.307$^{*}$} & \textbf{+0.151$^{**}$} & \textbf{-0.066} & +0.109 & -0.646 \\
openness low & -0.089 & +0.021 & \textbf{+0.068} & +0.029 & +0.056 \\
\bottomrule
\end{tabular}
\par\smallskip\footnotesize Differences from baseline (mean over seeds). $^{*}p<0.05$, $^{**}p<0.01$, $^{***}p<0.001$, paired $t$-test with Holm--Bonferroni correction within family $\times$ metric. Bold marks $|d_z|>0.8$. Baseline row shows absolute values (mean$\pm$sd).
\end{table}

\subsection{Effects hold across a wider set of topics}\label{sec:results-topic}

\begin{figure}[t]\centering
\includegraphics[width=\linewidth]{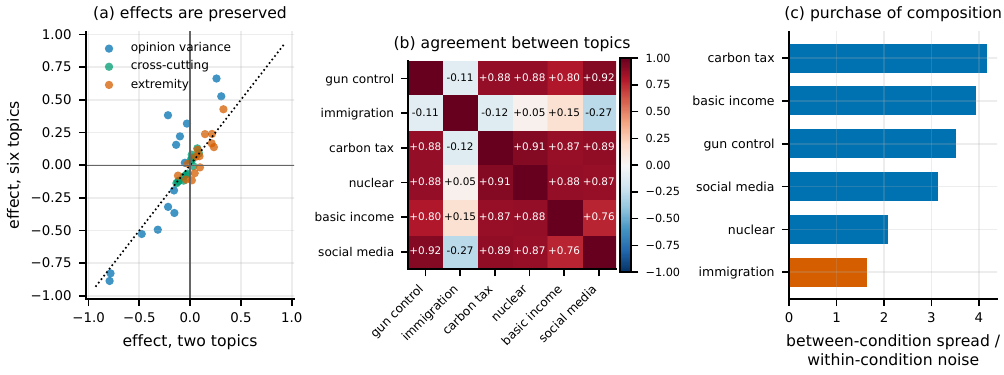}
\caption{The six-topic replication.
\textbf{(a)} Effect sizes estimated on six topics against the published two-topic estimates,
for every condition and three measures; the dotted line is equality.
\textbf{(b)} Rank agreement between each pair of topics across the fifteen conditions. Ten
of the fifteen pairs agree closely; the five exceptions all involve immigration.
\textbf{(c)} How far composition moves each topic, as the ratio of between-condition spread
to within-condition seed noise. Immigration is the topic composition moves least, which is
why its ordering across conditions is dominated by noise rather than opposed to the others.}
\label{fig:topics}
\end{figure}

Our polarization measures average over contested topics, and with only two of them any
averaged effect could be a property of that pair rather than of composition. In the
two-topic design the concern was concrete: the two topics disagreed, with a rank agreement
across conditions of $\rho = -0.52$. We therefore repeated the trait-level and heterogeneity
conditions with six contested topics rather than two, retaining the original pair so the two
designs are directly comparable. This adds 120 runs, at eight seeds per condition (Figure~\ref{fig:topics}).

\emph{Most effects hold, and the largest strengthen; five reverse sign.} Effect sizes estimated on six topics correlate
closely with the published two-topic estimates: $r = +0.90$ for opinion variance, $+0.93$ for
cross-cutting interaction, $+0.87$ for extremity and $+0.67$ for echo-chamber closure. The
principal effects are larger, not smaller, when six topics are averaged: the
human-calibrated condition moves opinion variance by $-0.887$ against $-0.792$ before, the
homogeneous condition by $-0.829$ against $-0.783$, high Agreeableness by $-0.527$ against
$-0.476$ and high Openness by $-0.494$ against $-0.318$. Sign agreement between the two
designs is 82\% for contrasts whose two-topic effect exceeded $0.10$ and 61\% for those
below it, so the disagreement is concentrated where effects are close to zero and their sign
is arbitrary.

\emph{Five effects reverse sign, and we name them.} Sixteen of the fifty-six
condition-by-measure combinations disagree in sign between the two designs. Eleven of those
have a two-topic effect below $0.10$ in absolute value, where the sign is not meaningfully
determined. Five do not, and these are genuine reversals rather than noise: lowered
Agreeableness on opinion variance ($-0.218$ on two topics, $+0.382$ on six) and on
echo-chamber closure ($+0.229$, $-0.123$); lowered Extraversion on the same two measures
($-0.137$, $+0.156$; $+0.162$, $-0.080$); and raised Neuroticism on echo-chamber closure
($+0.125$, $-0.122$). Four of the five involve a lowered trait, which is consistent with the
low-power account above: these were the contrasts the two-topic design estimated least
reliably. Where the two designs disagree we take the six-topic estimate as the better one,
and no claim in this article rests on any of the five.

\emph{The disagreement was one topic, not a general property.} With six topics we can
examine all fifteen pairs rather than one. Ten of the fifteen agree strongly, with a mean
rank agreement of $+0.87$ (range $+0.76$ to $+0.92$). Every one of the five exceptions
involves immigration, which agrees with the other topics at $-0.06$ on average (range
$-0.27$ to $+0.15$; Mann--Whitney $p = 0.001$ against the remaining pairs). Each of the
other five topics agrees with its peers at between $+0.63$ and $+0.72$.

The explanation appears to be sensitivity rather than opposition. Composition moves opinion
variance on immigration least of any topic: the spread of condition means is $1.13$ against
$2.53$ for gun control and $2.33$ for a carbon tax, and the ratio of between-condition spread
to within-condition seed noise is $1.6$ for immigration against a mean of $3.4$ for the
others. Immigration does not respond in the opposite direction; it responds weakly, so its
ordering across conditions is dominated by noise. The negative correlation we reported from
the two-topic design was therefore an artifact of pairing the most responsive topic with the
least responsive one.

\emph{Composition matters more than topic.} Decomposing the variance in opinion variance
across the six-topic runs, composition accounts for 53.2\% ($F = 77.0$, $p < 10^{-130}$) and
topic for 12.3\% ($F = 50.0$, $p < 10^{-43}$). Both matter and both are firmly established,
but composition accounts for more than four times as much.

We therefore report the topic-averaged effects as properties of composition rather than of
the topic set, while noting that topics differ in how much purchase composition has on them,
and that a study using only immigration would have found much weaker effects than one using
only gun control.

\subsection{Heterogeneity is the strongest compositional variable}\label{sec:results-e2}\label{sec:results-hetero}

\begin{figure}[t]\centering
\includegraphics[width=\linewidth]{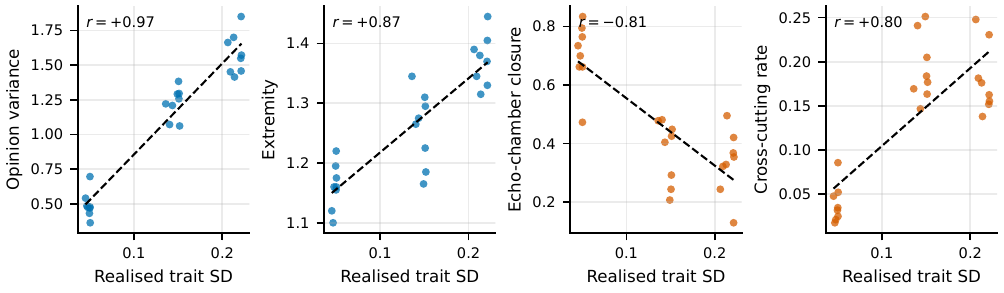}
\caption{Personality heterogeneity acts in opposite directions on the two faces of
polarization. Each point is one run of the heterogeneity conditions, in which all trait means
are held at the midpoint. Realised trait dispersion is positively associated with the
dispersion of opinion (variance, extremity; left two panels, blue) and negatively associated
with its segregation (echo-chamber closure; third panel, orange), while cross-cutting
interaction rises with it.}
\label{fig:hetero}
\end{figure}

E2 holds all trait means at the midpoint and varies only dispersion. Thirteen contrasts
survive correction (Table~\ref{tab:e2}), and the relationship with realised dispersion is
close to deterministic (Figure~\ref{fig:hetero}):

\begin{center}
\small
\begin{tabular}{lcc}
\toprule
Outcome & $r$ with realised trait SD & $p$ \\
\midrule
Opinion variance      & $+0.966$ & $2\times10^{-14}$ \\
Extremity             & $+0.870$ & $3\times10^{-8}$ \\
Cross-cutting rate    & $+0.795$ & $3\times10^{-6}$ \\
Echo-chamber closure  & $-0.810$ & $2\times10^{-6}$ \\
Collective accuracy   & $+0.774$ & $9\times10^{-6}$ \\
\bottomrule
\end{tabular}
\end{center}

The signs are the substantive finding. Personality diversity raises the \emph{dispersion} of
opinion and simultaneously lowers its \emph{segregation}: more varied societies hold more
varied views, talk across their divisions more, and cluster less. The homogeneous society is
not a moderate society but a consensual echo chamber, with opinion variance $0.783$ below
baseline ($d_z = -3.31$, $p < 0.001$), cross-cutting collapsed by $0.120$ ($d_z = -3.99$,
$p < 0.001$) and echo-chamber closure raised by $0.334$ ($d_z = 1.87$, $p = 0.003$).

\emph{The human-calibrated society occupies the least favourable region.} Constructing a
population from published IPIP-NEO-120 norms produced the most insular condition in the
study: the lowest cross-cutting rate ($-0.135$, $d_z = -3.04$, $p < 0.001$), the highest
echo-chamber closure ($+0.486$, $d_z = 3.59$, $p < 0.001$), the highest extremity ($+0.328$,
$d_z = 3.91$, $p < 0.001$), and opinion variance $0.792$ below baseline. It is also the only
condition with significant deficits on collective intelligence: collective accuracy $-0.275$
($d_z = -1.81$, $p = 0.006$), median relative error worse by $0.396$ ($p = 0.040$), and
estimate diversity reduced by $0.026$ ($p = 0.048$). This condition inherits the sampling
limitations of its normative source, discussed in Section~\ref{sec:discussion}; it should be
read as one plausible human-like composition, not as a population estimate.

\begin{table}[t]\centering
\scriptsize\setlength{\tabcolsep}{1.5pt}
\caption{Effects of trait heterogeneity, and of a human-calibrated composition (E2).}
\label{tab:e2}
\begin{tabular}{@{}lccccc@{}}
\toprule
Condition & Opinion var. & Extremity & Cross-cut & Echo closure & Coll. accuracy \\
\midrule
\textit{norm baseline} (baseline) & 1.263\,{\scriptsize$\pm$0.209} & 1.296\,{\scriptsize$\pm$0.077} & 0.165\,{\scriptsize$\pm$0.039} & 0.361\,{\scriptsize$\pm$0.163} & 0.106\,{\scriptsize$\pm$0.689} \\
\midrule
diverse & \textbf{+0.318$^{*}$} & \textbf{+0.096} & +0.018 & -0.056 & -0.048 \\
homog & \textbf{-0.772$^{***}$} & \textbf{-0.116$^{**}$} & \textbf{-0.123$^{***}$} & \textbf{+0.313$^{**}$} & \textbf{-0.735} \\
human norm & \textbf{-0.781$^{***}$} & \textbf{+0.332$^{***}$} & \textbf{-0.137$^{***}$} & \textbf{+0.466$^{***}$} & \textbf{-0.921$^{*}$} \\
mid & -0.041 & -0.018 & +0.030 & -0.017 & +0.072 \\
\bottomrule
\end{tabular}
\par\smallskip\footnotesize Differences from baseline (mean over seeds). $^{*}p<0.05$, $^{**}p<0.01$, $^{***}p<0.001$, paired $t$-test with Holm--Bonferroni correction within family $\times$ metric. Bold marks $|d_z|>0.8$. Baseline row shows absolute values (mean$\pm$sd).
\end{table}

\subsection{Agreeableness moderates the effect of Openness}\label{sec:results-surface}

Crossing Openness and Agreeableness reveals why single-trait accounts conflict.

A quadratic response surface fitted to opinion variance describes the grid well
($R^2 = 0.794$), but the unit of analysis matters for what can be inferred from it. Fitting
six parameters to nine cell means and computing $p$ values on the 72 underlying runs treats
within-cell simulation noise as replication and overstates the evidence. We therefore report
the interaction at the level at which the design actually varies. Refitting a reduced model
on the nine cell means gives $b(\mu_O \times \mu_A) = -3.86$ ($p = 0.030$, $\mathrm{df} = 5$);
a mixed model with cell as the grouping factor gives the same coefficient at $p = 0.0027$.

Four parameters on nine cells leaves little room, and the fit should be examined rather than
reported alone. In the primary model two cells exceed a Cook's distance of one, and although
the sign of the interaction is stable under leaving out any single cell, its significance is
not: $p < 0.05$ survives in three of the nine leave-one-out fits. The primary model's
estimate is therefore influence-driven and we do not rest the finding on it. The replication
models are what carry it: in Qwen2.5-14B no cell exceeds a Cook's distance of one and the
interaction remains significant in all nine leave-one-out fits, and in Qwen2.5-32B in seven
of nine. We report the crossover as established by replication across models rather than by
the fit in any one of them.
The interaction is therefore supported, but at conventional rather than extraordinary
strength, and we report it as such (Table~\ref{tab:e3surface}).

The interaction is a genuine crossover (Additional file~1). Among disagreeable
societies ($\mu_A = 0.2$), raising Openness increases opinion variance from $0.53$ to
$1.38$. Among agreeable societies ($\mu_A = 0.8$), the same manipulation \emph{decreases} it
from $0.97$ to $0.44$. Openness has no interpretable main effect on polarization; its sign
depends on the society in which it acts.

Cross-cutting interaction follows the same structure in the replication model, where the
interaction is significant at cell level ($b = -0.86$, $p < 0.05$), but not in the primary
model, where it is of the same sign and magnitude ($b = -0.49$) without reaching significance
on nine cells. We therefore claim the crossover for opinion variance and note it as
directionally consistent, not established, for cross-cutting. Eighteen individual cells of
the grid survive correction on the run-level contrasts. The jointly open and agreeable cell is the least polarized condition in the study
(variance $0.835$ below baseline, $d_z = -4.52$, $p < 0.001$) while also showing the largest
reduction in cross-cutting ($-0.125$, $p = 0.002$) and a large rise in echo-chamber closure
($+0.444$, $p = 0.002$): consensus purchased at the cost of contact.

\begin{table}[t]\centering
\scriptsize\setlength{\tabcolsep}{1.5pt}
\caption{Response-surface regression on the nine cell means of the Openness $\times$ Agreeableness grid.}
\label{tab:e3surface}
\begin{tabular}{@{}llccccc@{}}
\toprule
Model & Outcome & cells & $R^2$ & $\mu_O$ & $\mu_A$ & $\mu_O\!\times\!\mu_A$ \\
\midrule
Llama-3.1-8B & Opinion var. & 9 & 0.698 & +1.973$^{*}$ & +1.432 & -3.863$^{*}$ \\
Llama-3.1-8B & Cross-cut & 9 & 0.457 & +0.217 & +0.102 & -0.494 \\
Qwen2.5-7B & Opinion var. & 9 & 0.777 & +1.233 & +2.463 & -0.018 \\
Qwen2.5-7B & Cross-cut & 9 & 0.634 & +0.177 & +0.309 & -0.139 \\
Qwen2.5-14B & Opinion var. & 9 & 0.945 & +0.884 & +0.571 & -3.249$^{**}$ \\
Qwen2.5-14B & Cross-cut & 9 & 0.793 & +0.188 & +0.542$^{*}$ & -0.862$^{*}$ \\
Qwen2.5-32B & Opinion var. & 9 & 0.908 & +0.377 & +0.091 & -0.979$^{*}$ \\
Qwen2.5-32B & Cross-cut & 9 & 0.587 & +0.095 & +0.095 & -0.282 \\
\bottomrule
\end{tabular}
\par\smallskip\footnotesize Fitted on cell means, which is the level at which the design varies; fitting the full quadratic to nine cells and taking $p$ values from the underlying runs would treat within-cell simulation noise as replication. A mixed model with cell as the grouping factor gives the same interaction coefficients at $p = 0.003$, $p < 0.001$ and $p = 0.004$ for the 8, 14 and 32 billion parameter models. Only models passing induction validation are shown. $^{*}p<0.05$, $^{**}p<0.01$, $^{***}p<0.001$.
\end{table}

\subsection{Collective intelligence: one measure works, one fails}\label{sec:results-ci}

Of the two collective-intelligence measures specified in advance, post-discussion accuracy
discriminates strongly between conditions and the pre-to-post \emph{change} in accuracy does
not. The change measure shows no between-condition signal whatever (one-way $F = 0.35$,
$p = 0.995$), with between-condition variation roughly a fifth of within-condition variation.

The reason is substantive rather than technical. Private numeric estimates barely move across
the discussion window: on two of the six estimation items the mean absolute change in squared
log error is below $5\times10^{-4}$. Agents converse extensively and update their stated
opinions on contested topics, yet revise their private factual estimates almost not at all.
Composition therefore acts on collective accuracy through \emph{who the agents are} and how
their independent estimates aggregate, rather than through what deliberation does to them.
We report the change measure as a null and treat accuracy level as the primary
collective-intelligence outcome.

The hidden-profile task behaved as the classical paradigm predicts. Post-discussion accuracy
sat near $0.12$ across conditions: societies overwhelmingly selected the candidate favoured
by the information everyone already held, and discussion rarely surfaced enough unshared
information to overturn it. This replicates the classical failure in a new setting, but the
floor compresses the measure's ability to separate conditions, and only one grid cell
produced a corrected-significant difference. In the replication model the floor is
absolute: accuracy sat at or extremely near zero in every run, so the measure carries no
information there at all, and several contrasts involving it are undefined rather than
null (Additional file~1).

\subsection{Two further measures, reported for completeness}\label{sec:results-further}

Two of the six polarization measures defined in Section~\ref{sec:formulation} were computed
throughout but not reported above, and reporting only the four that discriminate would
misrepresent the instrument.

\emph{Assortativity discriminates nothing.} Numeric assortativity of opinion over the
follower graph has a mean of $-0.022$ and a standard deviation of $0.038$ across all
primary-model runs, and no condition moves it by more than $0.024$. Whatever structure
composition induces in these societies, it is not visible as opinion assortativity on the
network. We report this as a measure that failed rather than omitting it.

\emph{Bimodality moves but is not independent.} Sarle's coefficient rises under high
Agreeableness ($+0.121$) and the human-calibrated condition ($+0.122$) and falls under low
Agreeableness ($-0.091$), a pattern consistent with the dispersion results. It correlates
with extremity at $r = +0.55$ and with cross-cutting at $r = -0.49$, so it tracks the
measures already reported rather than adding a dimension. It is undefined where a society
converges completely, which is itself informative about those conditions.

\subsection{Cross-cutting contact is associated with collective accuracy}\label{sec:results-align}

The motivating hypothesis was a trade-off: that compositions reducing polarization would also
reduce collective intelligence. Across 258 primary-model runs the data do not show one, and
one association runs the other way. Establishing that this is not an arithmetic artifact
requires care, and we set out that case before drawing any conclusion from it.

\begin{figure}[t]\centering
\includegraphics[width=\linewidth]{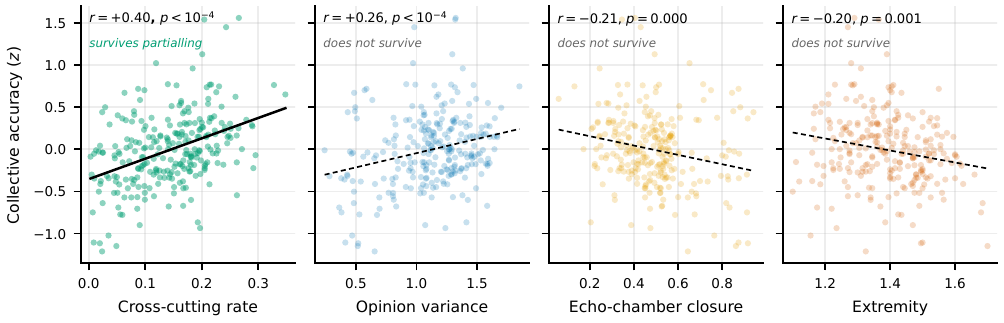}
\caption{Polarization measures against collective accuracy across 258 primary-model runs.
Only cross-cutting interaction retains its association once the components of the
diversity-prediction identity are partialled out; the remaining three are shown faded, with
dashed fits, because their unadjusted associations do not survive that adjustment.}
\label{fig:alignment}
\end{figure}

Cross-cutting interaction is positively associated with collective accuracy
($r = +0.41$, $p < 10^{-4}$) and opinion variance likewise ($r = +0.27$), while echo-chamber
closure ($r = -0.23$) and extremity ($r = -0.20$) are negatively associated with it
(Figure~\ref{fig:alignment}).

\emph{These correlations have an obvious mechanical explanation, which must be ruled out.}
Collective accuracy is measured by the error of the averaged estimate, and the
diversity-prediction decomposition states that this error equals mean individual error minus
the diversity of estimates. A composition producing more varied agents will therefore reduce
collective error \emph{by construction}, without anything social having occurred. Since
composition also drives the polarization measures, any association between them could be
common-cause rather than substantive --- particularly given our own finding
(Section~\ref{sec:results-ci}) that agents barely revise private estimates, which removes the
deliberative pathway that would otherwise explain it.

We therefore partial the association on the two components of the decomposition and on
realised trait dispersion, singly and jointly:

\begin{center}
\small
\begin{tabular}{@{}lcc@{}}
\toprule
Cross-cutting rate vs collective accuracy & $r$ & $p$ \\
\midrule
unadjusted                                         & $+0.409$ & $<10^{-4}$ \\
controlling estimate diversity                     & $+0.426$ & $<10^{-4}$ \\
controlling mean individual accuracy               & $+0.301$ & $<10^{-4}$ \\
controlling realised trait dispersion              & $+0.363$ & $<10^{-4}$ \\
controlling all three jointly                      & $+0.240$ & $0.0001$ \\
\bottomrule
\end{tabular}
\end{center}

The association is attenuated but not eliminated: more than half of the raw correlation
survives the removal of every component that could produce it arithmetically. Applying the
same treatment to the other three measures removes them entirely. Opinion variance falls from
$+0.268$ to $+0.071$ ($p = 0.26$), echo-chamber closure from $-0.230$ to $-0.064$
($p = 0.31$), and extremity from $-0.200$ to $-0.120$ ($p = 0.055$). We withdraw all three as
evidence: their unadjusted associations with accuracy are attributable to the aggregation
identity. \emph{Cross-cutting interaction is the only polarization measure whose association
with collective accuracy survives.}

A stronger test is whether cross-cutting predicts accuracy \emph{within} a condition, where
composition is constant and a common cause is impossible by construction. Here the evidence
is suggestive rather than conclusive: the mean within-condition correlation is $+0.107$, and
16 of 24 conditions are positive, but this does not reach significance ($t = 1.40$,
$p = 0.18$) at eight runs per condition.

We therefore state the finding at the strength the evidence supports. There is no trade-off:
no polarization measure predicts poorer collective performance once the aggregation identity
is removed. Beyond that, a single positive association --- between cross-cutting interaction
and collective accuracy --- is robust to the arithmetic explanation, but rests on variation
between compositions rather than within them. Whether contact itself improves accuracy, or
whether compositions that generate contact happen also to generate better-calibrated agents,
cannot be settled by this design. Separating them would require manipulating contact
independently of composition, for instance through the recommender, which
Section~\ref{sec:discussion} identifies as the natural next experiment.

One further qualification. The hidden-profile measure is unrelated to every polarization
measure, and we state that as an equivalence rather than as a failure to reject. Against an
equivalence bound of $|r| = 0.20$, a small effect by conventional standards, two one-sided
tests reject the presence of an association of at least that size for all four measures
(opinion variance $r = -0.02$, $p = 0.0007$; extremity $r = +0.06$, $p = 0.005$;
cross-cutting $r = -0.09$, $p = 0.019$; echo closure $r = +0.05$, $p = 0.003$). The
association, if any, is smaller than a small effect. The result therefore rests on the
estimation measure alone. Given the floor effect described above, we read this as a limitation of the
hidden-profile task in this setting rather than as evidence against the association.

\subsection{Robustness to design perturbation}\label{sec:results-robust}

Each of the two anchor conditions was compared against the baseline under the same
perturbation --- the baseline is the comparison, not a third anchor --- across
two alternative network topologies, halved and increased activation, doubled memory, an
enlarged feed, raised temperature, and a change of model. Table~\ref{tab:e5audit} reports the
results as sign stability; the value for each individual perturbation is given in
Additional file~1.

Every polarization effect tested is stable. High Openness lowers opinion variance in all nine
cells (range $-0.32$ to $-0.07$), lowers cross-cutting in all nine ($-0.14$ to $-0.05$), and
raises echo-chamber closure in all nine ($+0.02$ to $+0.20$). High Neuroticism raises opinion
variance in all nine ($+0.08$ to $+0.29$), raises extremity in all nine ($+0.21$ to $+0.36$),
and raises closure in all nine ($+0.11$ to $+0.25$).

The collective-intelligence effects are not stable: accuracy holds its sign in six of nine
cells for one anchor and seven of nine for the other, and hidden-profile accuracy in two to
five of nine. These are reported as exploratory. Where an anchor's reference effect is near
zero --- Neuroticism on cross-cutting, at $+0.016$ --- the sign of a null is arbitrary and low
stability carries no information.

One polarization effect is not fully stable: high Openness on extremity holds in eight of
nine cells, and the single exception is the change of model family, which the next section
takes up.

Increasing society size from 100 to 200 agents left the baseline broadly unchanged (extremity
$1.276$ versus $1.280$; hidden-profile accuracy $0.106$ versus $0.130$), with somewhat higher
opinion variance ($1.455$ versus $1.275$) and lower echo-chamber closure ($0.268$ versus
$0.369$). Larger societies are slightly more dispersed and less closed, and no conclusion
depends on scale.

\begin{table}[t]\centering
\scriptsize\setlength{\tabcolsep}{3pt}
\caption{Robustness audit. Each anchor condition is compared against the baseline \emph{under the same perturbation}, across two alternative network topologies, halved and increased activation, doubled memory, an enlarged feed, raised temperature, and a change of model family. Stability is the number of cells, out of nine, in which the effect preserves the sign of its unperturbed reference value.}
\label{tab:e5audit}
\begin{tabular}{@{}llccccc@{}}
\toprule
Anchor & Metric & Reference & Min & Max & Stability & Verdict \\
\midrule
openness high & Opinion var. & -0.306 & -0.306 & -0.067 & 9/9 & \textbf{robust} \\
openness high & Extremity & +0.131 & -0.068 & +0.237 & 8/9 & exploratory \\
openness high & Cross-cut rate & -0.068 & -0.144 & -0.045 & 9/9 & \textbf{robust} \\
openness high & Echo closure & +0.138 & +0.020 & +0.197 & 9/9 & \textbf{robust} \\
openness high & Collective acc. & -0.521 & -0.521 & +0.167 & 5/9 & exploratory \\
openness high & Hidden profile & -0.017 & -0.022 & +0.020 & 5/9 & exploratory \\
neuroticism high & Opinion var. & +0.272 & +0.076 & +0.291 & 9/9 & \textbf{robust} \\
neuroticism high & Extremity & +0.218 & +0.207 & +0.355 & 9/9 & \textbf{robust} \\
neuroticism high & Cross-cut rate & +0.011 & -0.041 & +0.068 & 3/9 & exploratory \\
neuroticism high & Echo closure & +0.134 & +0.111 & +0.250 & 9/9 & \textbf{robust} \\
neuroticism high & Collective acc. & -0.192 & -0.192 & +0.450 & 3/9 & exploratory \\
neuroticism high & Hidden profile & +0.015 & -0.027 & +0.015 & 2/9 & exploratory \\
\bottomrule
\end{tabular}
\par\smallskip\footnotesize Per-perturbation values are released in \texttt{results/audit\_sign\_stability.csv}.
\end{table}

\subsection{Replication across model scale and family}\label{sec:results-crossmodel}

\begin{figure}[t]\centering
\includegraphics[width=\linewidth]{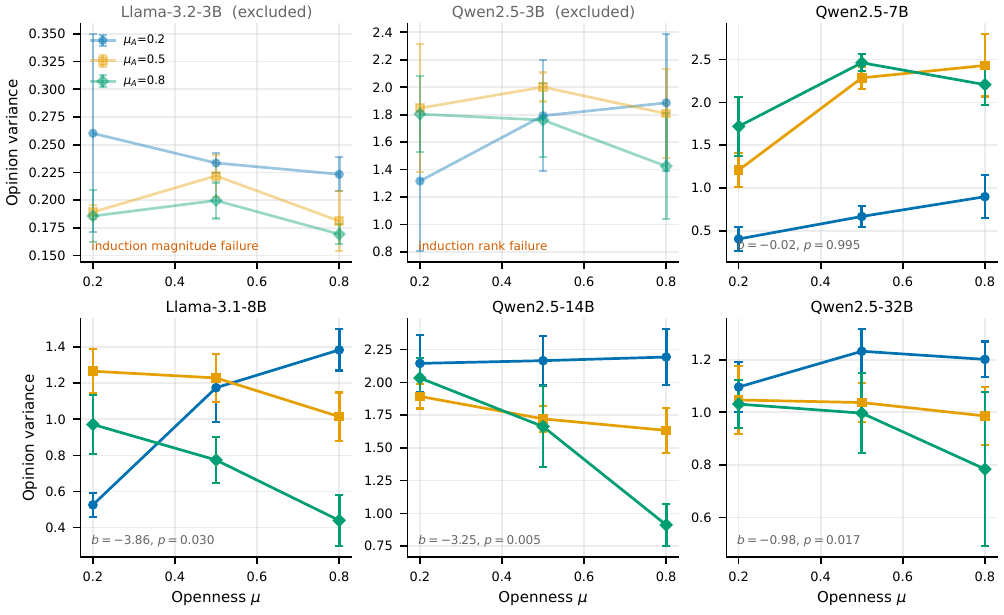}
\caption{Response surfaces for every model tested. Opinion variance against mean Openness at
each level of mean Agreeableness, with error bars giving standard deviations over seeds.
The two faded panels are models excluded by induction validation; note that the vertical
range in the first spans $0.2$ units against $0.9$ for Llama-3.1-8B, so its apparent
structure is smaller than it looks. Among the four admitted models the interaction appears in
three, with Qwen2.5-7B showing additive rather than crossed effects.}
\label{fig:sixmodel}
\end{figure}

The response surface was repeated on five further models, spanning 3 to 32 billion parameters
and two families. Induction was validated separately in each, on the same nine
configurations, before any of their results were interpreted (Table~\ref{tab:induction},
Figure~\ref{fig:sixmodel}).

\emph{Two models are excluded, for different reasons.} Qwen2.5-3B fails the pre-specified
rank criterion on Agreeableness ($\rho = 0.48$ against a threshold of $0.60$), so the
ordering of its configurations on that trait is not reliable. Llama-3.2-3B passes on rank
($\rho = 0.85$ and $0.96$) but fails on magnitude: the full manipulation moves its measured
scores by $0.35$ and $0.71$ points, against $2.22$ and $3.06$ in the primary model. Its
behaviour agrees --- the range of its condition means in opinion variance is $0.09$, against
$0.45$ to $2.06$ in every admitted model. In that model the manipulation was too weak to test
anything, and its flat surface is a statement about induction rather than about composition.
We note that excluding it removes a model that would otherwise appear to contradict the
interaction, and for that reason report its surface in Additional file~1.

\emph{Among the four admitted models, the interaction appears in three.} It is negative and
significant in the primary model ($b = -3.86$, $p = 0.030$), at 14 billion parameters
($b = -3.25$, $p = 0.005$) and at 32 billion ($b = -0.98$, $p = 0.017$); mixed models with
cell as the grouping factor agree ($p = 0.003$, $p < 0.001$, $p = 0.004$). Since these span
two families and a fourfold range of parameter count, the moderation is not an artifact of one
training pipeline or of one model size.

\emph{One admitted model behaves differently, and it is not a measurement failure.}
Qwen2.5-7B shows no interaction at all ($b = -0.02$, $p = 0.995$). Its Openness slope is
$+0.49$ at low Agreeableness and $+0.49$ at high: the two traits act additively, each raising
opinion variance independently of the other. This is not weak induction. It passes both
criteria comfortably ($\Delta = 1.13$ and $1.93$), and its range of condition means is
$2.06$, the largest of any model we ran --- it responds to composition more strongly than the
models in which the crossover appears, but in a different form. We report it as a substantive
exception, and we characterised it further by repeating the trait-level and heterogeneity
experiments on that model.

\begin{figure}[t]\centering
\includegraphics[width=0.62\linewidth]{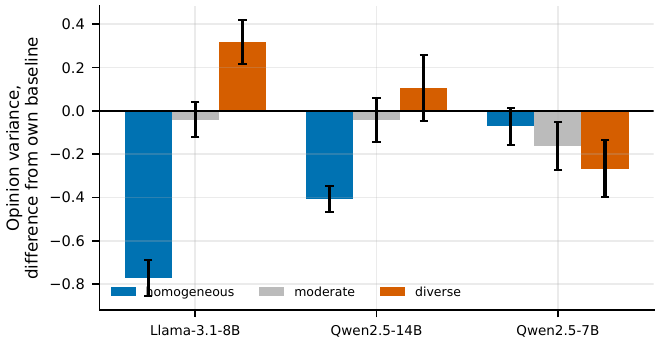}
\caption{The heterogeneity effect in three models. Bars give the difference in opinion
variance from each model's own baseline, with standard errors over seeds. The ordering that
defines the effect --- homogeneous societies below baseline, diverse societies above --- holds
in the primary model and in Qwen2.5-14B and is absent in Qwen2.5-7B, the model that also
shows no Openness by Agreeableness interaction.}
\label{fig:heteromodels}
\end{figure}

\emph{The exception is specific to traits acting jointly.} On trait levels, Qwen2.5-7B is
unremarkable: its sign agreement with the primary model across the forty trait-level
contrasts is 23/40, marginally higher than the 22/40 of the replication model that does show
the interaction. Traits taken one at a time behave in it much as they behave anywhere. On
heterogeneity, however, it departs sharply. In the primary model and in Qwen2.5-14B,
homogeneous societies fall well below baseline in opinion variance and diverse societies rise
above it ($-0.78$ and $+0.31$; $-0.41$ and $+0.11$). In Qwen2.5-7B neither holds: the
homogeneous condition is flat ($-0.07$) and the diverse condition moves in the wrong
direction ($-0.27$). The dose-response relation that underpins Section~\ref{sec:results-hetero}
is correspondingly absent: realised trait dispersion correlates with opinion variance at
$r = +0.97$ in the primary model and $r = +0.75$ in Qwen2.5-14B, but at $r = -0.33$
($p = 0.11$) in Qwen2.5-7B (Figure~\ref{fig:heteromodels}).

The pattern is therefore coherent rather than arbitrary. Both of our compositional findings
that require traits to act \emph{jointly} --- their interaction, and their dispersion ---
fail in this model, while the finding concerning traits acting \emph{singly} does not. In
Qwen2.5-7B the traits appear to operate independently and additively, and neither their
combination nor their spread has purchase on the society's behaviour. Why one model should
lack the joint structure that the others exhibit is not something this design can answer, and
we note that this is a single model; but the failure is structured, and a structured failure
is a more useful thing to report than an unexplained one.

\emph{There is no systematic relationship with scale.} The interaction is present at 8, 14
and 32 billion parameters and absent at 7; the range of condition means across the four
admitted models is $2.06$, $0.94$, $1.28$ and $0.45$ in ascending order of size. Neither the
interaction nor overall sensitivity to composition is monotone in parameter count. We
mention this because an earlier version of this analysis, based on three models, suggested an
attenuation with scale that the additional models did not support.

\begin{table}[t]\centering
\scriptsize
\setlength{\tabcolsep}{3.5pt}
\caption{Induction validation and response-surface shape, by model.}
\label{tab:induction}
\begin{tabular}{@{}p{2.3cm}p{1.5cm}cccccp{2.4cm}@{}}
\toprule
Model & Family & \multicolumn{2}{c}{Openness} & \multicolumn{2}{c}{Agreeableness} & Range & Surface \\
\cmidrule(lr){3-4}\cmidrule(lr){5-6}
 & & $\rho$ & $\Delta$ & $\rho$ & $\Delta$ & of means & \\
\midrule
\textit{Llama-3.2-3B} & Llama-3.2 & 0.85 & \textit{0.35} & \textit{0.96} & 0.71 & 0.09 & \textit{excluded, magnitude} \\
\textit{Qwen2.5-3B} & Qwen2.5 & 0.86 & \textit{1.32} & \textit{0.48} & 1.21 & 0.68 & \textit{excluded, rank} \\
Qwen2.5-7B & Qwen2.5 & 0.95 & 1.13 & 0.95 & 1.93 & 2.06 & additive \\
Llama-3.1-8B & Llama-3.1 & 0.95 & 2.22 & 0.95 & 3.06 & 0.94 & crossover \\
Qwen2.5-14B & Qwen2.5 & 0.97 & 1.39 & 0.95 & 2.17 & 1.28 & crossover \\
Qwen2.5-32B & Qwen2.5 & 0.95 & 2.31 & 0.95 & 1.84 & 0.45 & crossover \\
\bottomrule
\end{tabular}
\par\scriptsize
\setlength{\tabcolsep}{3.5pt}skip\footnotesize $\rho$ is the Spearman correlation between targeted and measured trait level across the nine grid configurations; $\Delta$ is the difference in mean measured score between the highest and lowest targeted level, on the instrument's 1--5 scale. A model is admitted only if $\rho \geq 0.60$ and $\Delta \geq 1.00$ for both manipulated traits. ``Range of means'' is the spread of condition means in opinion variance, the behavioural counterpart of $\Delta$. Italicised rows are excluded; both are reported in Additional file~1.
\end{table}

\emph{Trait-level main effects largely do not replicate.} Sign agreement across the ten
trait-level conditions is shown in Table~\ref{tab:crossmodel}; it is 6/10 for opinion variance, 6/10 for cross-cutting, 6/10 for
echo-chamber closure and 4/10 for extremity (Table~\ref{tab:crossmodel}). At ten conditions
these figures are close to what chance would produce, and we do not claim them as
replication. Among individual effects, Agreeableness on cross-cutting reproduces almost
exactly ($-0.111$ in the primary model, $-0.106$ in the replication), and the direction of
the Agreeableness, Openness and Neuroticism effects on opinion variance and echo-chamber
closure is preserved. Extremity does not replicate: its sign reverses for most conditions.

Two independent observations explain the extremity failure. The robustness audit had already
flagged it as the one polarization measure whose sign flipped under a change of model, before
the full replication was run. And the artifact control tells us why: in the replication model,
filler-topic variance is strongly associated with contested-topic extremity ($r = +0.367$,
$p < 10^{-4}$, $n = 129$), the signature of response-style variation that is absent in the
primary model ($r = -0.061$, $p = 0.33$). Extremity there is contaminated by a general
tendency to answer scales more extremely, and is not a trustworthy polarization measure.

Absolute levels differ substantially between models --- baseline opinion variance $1.76$
versus $1.27$, baseline echo-chamber closure $0.06$ versus $0.37$, hidden-profile accuracy
$0.001$ versus $0.122$ --- so these simulations should not be read as estimating population
quantities. What transfers is the moderation, not the magnitudes and not the main effects.

\begin{table}[t]\centering
\scriptsize
\setlength{\tabcolsep}{3pt}
\caption{Cross-model replication of trait-level effects. Each cell is the difference from that model's own baseline.}
\label{tab:crossmodel}
\begin{tabular}{@{}lccccccccc@{}}
\toprule
 & \multicolumn{3}{c}{Opinion var.} & \multicolumn{3}{c}{Extremity} & \multicolumn{3}{c}{Cross-cut} \\
Condition & L & Q14 & Q7 & L & Q14 & Q7 & L & Q14 & Q7 \\
\midrule
agreeableness high & -0.46 & \textbf{-0.20} & +0.30 & +0.20 & -0.05 & -0.03 & -0.12 & \textbf{-0.11} & \textbf{-0.04} \\
agreeableness low & -0.21 & +0.37 & \textbf{-1.53} & +0.08 & \textbf{+0.15} & \textbf{+0.27} & +0.00 & -0.09 & -0.22 \\
conscientiousness high & -0.15 & +0.03 & +0.10 & +0.20 & -0.03 & \textbf{+0.02} & -0.07 & +0.02 & \textbf{-0.03} \\
conscientiousness low & -0.02 & \textbf{-0.06} & \textbf{-0.78} & -0.06 & \textbf{-0.03} & +0.08 & +0.08 & -0.03 & -0.11 \\
extraversion high & -0.20 & \textbf{-0.03} & +0.23 & +0.06 & -0.05 & \textbf{+0.02} & -0.04 & \textbf{-0.10} & \textbf{-0.03} \\
extraversion low & -0.13 & +0.11 & \textbf{-0.53} & +0.03 & \textbf{+0.05} & \textbf{+0.10} & +0.01 & \textbf{+0.08} & -0.12 \\
neuroticism high & +0.27 & \textbf{+0.07} & -1.78 & +0.22 & \textbf{+0.16} & \textbf{+0.36} & +0.01 & \textbf{+0.03} & -0.26 \\
neuroticism low & -0.14 & \textbf{-0.14} & +0.21 & +0.04 & -0.08 & -0.07 & -0.03 & \textbf{-0.07} & \textbf{-0.01} \\
openness high & -0.31 & \textbf{-0.28} & +0.01 & +0.13 & -0.14 & \textbf{+0.01} & -0.07 & \textbf{-0.01} & \textbf{-0.03} \\
openness low & -0.09 & +0.01 & \textbf{-0.94} & +0.00 & -0.01 & \textbf{+0.06} & +0.07 & -0.02 & -0.13 \\
\midrule
Sign agreement & \multicolumn{3}{c}{10/20} & \multicolumn{3}{c}{11/20} & \multicolumn{3}{c}{11/20} \\
\bottomrule
\end{tabular}
\par\smallskip\footnotesize L = Llama-3.1-8B, Q14 = Qwen2.5-14B, Q7 = Qwen2.5-7B. Bold marks a replication cell whose sign agrees with the primary model. Only models passing induction validation (Table~\ref{tab:induction}) are included.
\end{table}

\section{Discussion}\label{sec:discussion}

\subsection{What the results say}

Three findings stand out, and they are of different kinds.

The first is structural. \emph{Polarization is not one thing.} Across every experiment, the
dispersion of opinion and its segregation into camps moved independently and sometimes in
opposite directions. Personality heterogeneity raised dispersion while lowering segregation;
high Agreeableness lowered dispersion while raising segregation. A society can be
homogeneous and closed, or varied and open, and the two axes are not redundant. Studies that
report a single polarization statistic can therefore reach opposite conclusions about the
same system depending on which statistic they chose, without either being wrong.

The second is the interaction. \emph{Agreeableness determines the sign of Openness.} In
disagreeable societies, raising Openness increased opinion variance; in agreeable ones it
decreased it, and the crossover replicated in a second model family with a near-identical
interaction coefficient. This offers a concrete explanation for why single-trait findings in
this literature conflict: a study that manipulates Openness in a population that happens to
be agreeable will report the opposite effect to one that manipulates it in a population that
is not. Composition is not a nuisance parameter to be randomised away; it is the variable
that determines what the other variables do.

The third is the one we did not expect, and it is the one we state most cautiously.
\emph{We find no trade-off, and one positive association that survives scrutiny.} We designed
this study around a hypothesised trade-off and the data do not show one: no polarization
measure predicts poorer collective performance. Cross-cutting interaction predicts greater
collective accuracy, and that association survives partialling on both components of the
diversity-prediction identity and on realised trait dispersion. It is the only one of the four
polarization measures that does: the unadjusted associations of opinion variance, extremity
and echo-chamber closure with accuracy are attributable to the aggregation identity and we
withdraw them. That the finding narrows to a single measure under adjustment is, we think,
the correct outcome rather than a disappointing one --- it is the measure that describes
contact between disagreeing agents, which is the one a deliberative account would single out.
It nonetheless rests on variation between compositions rather than within them, and the
within-condition test is suggestive but not significant. Whether contact improves accuracy,
or whether compositions that generate contact also generate better-calibrated agents, this
design cannot separate. If the association is causal, it removes a supposed dilemma from
platform design; we report it as a robust association and not more.

The human-calibrated condition sharpens the point. Built from published population norms,
it landed in the most insular region of the design space --- lowest cross-cutting, highest
closure, highest extremity --- and was the only condition with significant deficits on
collective intelligence. Read cautiously, and subject to the sampling limitations discussed
below, that is a discouraging result: the composition closest to a real population is not a
comfortable place to be.

\subsection{Relation to prior work}

Our polarization findings extend the closest prior study in this journal. Cau et al.\
\cite{cau2025selective} report that uniform LLM populations converge toward agreement
regardless of initial conditions, and identify the absence of agent personalities and of
network structure as the two limitations of their design. Introducing both changes the
picture: convergence is not a general property of LLM populations but a property of
\emph{particular compositions}. Our high-Agreeableness societies converge in the way they
describe, while our high-Neuroticism societies do not converge at all, sustaining two stable
camps for the length of the simulation. Composition determines whether the convergence they
observe occurs.

Our collective-intelligence result speaks to an unsettled debate. Lorenz and colleagues
showed experimentally that social influence can undermine the wisdom of crowds by narrowing
diversity without improving accuracy \cite{lorenz2011social}. Becker and colleagues
subsequently showed that in decentralised networks social influence can \emph{improve}
group estimates \cite{becker2017network}, and Navajas and colleagues showed that
deliberation within small independent groups outperforms aggregating a large crowd
\cite{navajas2018aggregated}. Our findings sit closer to the latter: what predicted accuracy here was the breadth of
contact rather than isolation from influence. The mechanism, however, cannot be deliberative
in our system, since private estimates barely move; and the association survives the
arithmetic explanation without being demonstrable within a fixed composition. It is best read
as a hypothesis this design generates rather than one it settles.

The mechanism, however, differs from the deliberative one. Agents in our societies barely
revised their private numeric estimates through discussion, so the association between
cross-cutting contact and accuracy is unlikely to arise from belief updating during
deliberation. The more parsimonious reading is that compositions producing more cross-cutting
contact also produce populations whose independent estimates aggregate better --- an
aggregation effect rather than a deliberation effect, closer to the diversity-prediction
account \cite{page2008difference} than to a model of persuasion. Distinguishing these two
pathways would require manipulating contact independently of composition, which our design
does not do.

On the hidden-profile task we reproduce the classical failure \cite{stasser1985pooling}:
societies systematically chose the option favoured by information all members already held.
That the failure survives transplantation into a hundred-agent social network with an
algorithmic feed is a modest but genuine extension of a finding established in small
face-to-face groups.

Methodologically, the study is a response to recent arguments that validation is the central
unsolved problem in generative social simulation \cite{larooij2026validation}. Our answer is
not a new validation theory but a set of controls that can be run cheaply and reported
plainly: a pre-specified induction gate, a neutral filler topic to detect response-style
artifacts, exact reconstruction of realised trait distributions to bound a sampling
confound, unplanned internal replications, a sign-stability audit across design
perturbations, and a full replication in a second model family.

\subsection{The filler-topic control earned its place}

The value of such controls is illustrated by the extremity measure. In the primary model,
extremity behaved as a polarization measure should, and four conditions moved it
significantly; in the replication model, its sign reversed for most conditions. Absent a
control, two interpretations would remain open: that the effect is model-specific, or that
one model's measurement is unreliable.

The filler topic settles it. In the replication model, variance on a neutral topic about
pizza toppings correlates at $r = +0.367$ with extremity on gun control and immigration; in
the primary model the correlation is absent. The replication model exhibits general
response-style variation across conditions, so its extremity scores partly measure how
emphatically agents answer scales rather than how polarized they are. Extremity is therefore
not a trustworthy polarization measure in that model, and we do not treat its non-replication
as evidence against the underlying phenomenon.

A neutral control topic is inexpensive, requiring one additional probe per round, and
distinguishes a substantive effect from a prompt-sensitivity artifact. We suggest it be
adopted routinely in personality-conditioned simulation.

\subsection{Limitations}\label{sec:limitations}

\emph{These are simulated societies.} Nothing here establishes that human societies with
these compositions behave this way. The agents are language models role-playing personas;
their opinions are elicited responses, not beliefs. The appropriate reading is that we have
characterised the behaviour of a class of simulation, and generated hypotheses that human
studies could test.

\emph{Induction validity is questionnaire validity.} Our induction gate produced a
correlation of $0.93$ between targeted and measured traits --- higher than reported for much
larger instruction-tuned models. We do not take this as evidence of unusually good
personality induction. A model whose system prompt states that it is highly extraverted can
answer an extraversion questionnaire correctly by reading its own instructions, without any
deeper behavioural commitment. Questionnaire validity of this kind is necessary but far from
sufficient, and the gap between it and behavioural validity remains open.

\emph{Behavioural trait measurement is weak.} We logged expressed personality from agents'
posts as a continuous manipulation check, but the scorer available for the full dataset was
a lexical-marker proxy, which detects the vocabulary it was built around and is close to
circular. We therefore do not report expressed-trait drift as a finding. A trained
classifier was prepared but could not be applied retroactively without re-running the
programme with mixed instrumentation, which would have been worse than reporting the
limitation.

\emph{One collective-intelligence measure failed and another floored.} The change-in-accuracy
measure carried no between-condition signal, and the hidden-profile task sat near a floor of
$0.12$, compressing its ability to separate conditions. The alignment result therefore rests
on a single measure. Replication with a task calibrated to mid-range difficulty is the
obvious next step.

\emph{The human-calibrated condition is not human-representative.} Its normative statistics
come from a large but self-selected online sample whose own authors note is not nationally
representative and is likely elevated on Openness- and Altruism-related facets
\cite{kajonius2019structure} --- the very direction that drives the insular-consensus result.
The condition should be read as one plausible human-like composition.

\emph{Two structural dependencies.} Echo-chamber closure is partly mechanical when opinion
variance is very low, since almost every tie is then within-camp; we never interpret it
alone. And the filler control compares variance on the filler topic against extremity on the
contested topics, which are different statistics --- an imperfectly matched comparison,
though the direction of the primary-model result makes an artifact explanation unlikely.

\emph{The alignment result is an association between compositions, not a demonstrated
mechanism.} Cross-cutting contact is associated with collective accuracy after the components of the
diversity-prediction identity are partialled out, so it is not simply the arithmetic of
averaging more varied estimates. But the within-condition test, which would isolate the
effect from composition entirely, is not significant at eight runs per condition, and private
estimates barely move during discussion, which rules out the deliberative mechanism that
would most naturally explain it. Manipulating contact independently of composition ---
through the recommender rather than through personality --- is the experiment that would
settle it.

\emph{Replication is of the interaction, not of the main effects, and rests on three
models.} Sign agreement across the ten trait-level conditions is close to chance, and only
the Openness by Agreeableness moderation reproduces convincingly beyond the primary model.
That moderation now holds in three models spanning 8 to 32 billion parameters and two
families, which rules out its being an artifact of one developer's training pipeline or of
one model size. It does not establish that it holds for language models in general. Three
models are a small sample of a large and fast-moving space; the design is unbalanced, with
only one family represented at two sizes, so scale and family are partially confounded; and
all three are open-weight models in the 8--32 billion range, which excludes both much
smaller models and the largest proprietary systems. The attenuation we observe with scale
is the clearest signal that model choice matters to the magnitude of these effects even
where it does not change their direction, and it is the reason we state the cross-model
result as a limit on specificity rather than as a claim of generality. The released code
makes adding a further model a single command, and we regard broadening this comparison as
the most straightforward extension of the work.

\emph{Topics differ in how much composition can move them.} The two-topic design suggested
that composition acts on each topic differently, since the two disagreed. Extending to six
topics (Section~\ref{sec:results-topic}) shows this was a property of one topic rather than
of topics in general: five of the six agree closely with one another, and only immigration
stands apart, because composition has markedly less purchase on it. The topic-averaged
effects therefore describe composition rather than the topic set. What remains true, and
worth stating, is that the size of a composition effect depends on the question being
discussed: the same manipulation moves opinion variance on gun control more than twice as far
as on immigration. A study using a single topic could therefore report an effect much larger
or much smaller than ours, and topic sensitivity deserves to be reported as a variable in its
own right rather than averaged away. Why immigration resists composition is not something
this design can answer.

\emph{Scope.} One recommender family, thirty rounds, and societies of one hundred to two
hundred agents. The audit shows conclusions are stable across the design perturbations we
tested, but longer horizons and other feed algorithms remain untested.

\subsection{Implications and next steps}

For researchers building agent societies, the practical implication is that personality
composition should be reported and, where possible, controlled. A simulation that assigns
personas without describing their distribution has left a variable free that our results show
can reverse the sign of other effects.

For platform design, the alignment result is the substantive one: on these measures nothing
had to be traded away to reduce segregation. The natural follow-up is to manipulate feed
algorithms and personality composition together, since our design varies composition while
holding the recommender fixed, and the interaction between them is untested.

For methodology, the low-cost controls may prove more durable than the specific findings. The filler topic caught a model-specific artifact. The reconstruction of realised
trait distributions bounded a sampling confound at roughly one twelfth of the effect. The
unplanned internal replications gave a reliability estimate for free. And the sign-stability
audit separated six robust polarization effects from a set of collective-intelligence effects
that do not survive perturbation --- a separation we would not otherwise have known to make.

\section{Conclusion}\label{sec:conclusion}

We set out to determine whether the Big Five composition of a simulated online society shapes
polarization and collective intelligence at the same time, and whether the two trade off
against one another. Across 991 simulations spanning trait levels, trait heterogeneity, a
response surface, a robustness audit and a replication in a second model family, the answer
to the first question is yes and to the second is no.

Composition matters, and it matters in a structured way. Personality heterogeneity is the
compositional variable with the largest measured effects, and it acts in opposite
directions on the
two faces of polarization: more varied societies hold more dispersed opinions while being
less segregated into camps. Homogeneous societies are not moderate; they are consensual echo
chambers. Trait levels matter too, and their effects are not additive: Agreeableness
determines whether raising Openness increases or decreases polarization, an interaction that replicated with a near-identical coefficient in a second model family,
although the trait-level main effects did not.

The trade-off we designed the study to measure does not appear: no polarization measure
predicts poorer collective performance. Cross-cutting interaction predicts greater collective accuracy, and is the
only one of four polarization measures whose association survives removal of the aggregation
identity that could otherwise explain it, though it rests on comparisons between compositions
rather than within them. If it proves causal, it would remove a supposed dilemma from platform design.

The controls also merit attention. A neutral filler topic, requiring one additional probe
per round, revealed that a measure's failure to replicate across model families
was a response-style artifact rather than a substantive disagreement. Reconstructing realised
trait distributions bounded a sampling confound at roughly one twelfth of the effect it might
have explained. A sign-stability audit separated six polarization effects that survive every
design perturbation from a set of collective-intelligence effects that do not. In a field where reported emergent
phenomena have been shown to vanish under scrutiny, controls of this kind are inexpensive
and we encourage their routine adoption.

The clearest next step follows from what we did not vary. Composition was manipulated while
the recommender was held fixed; the interaction between personality composition and feed
design is untested, and is where the practical leverage most plausibly lies. Equally open is
the gap between questionnaire and behavioural validity of induced personality, which our own
induction gate measures only on the questionnaire side. All code, configurations and
run-level data are released so that both questions can be taken up directly from where this
study leaves off.

\backmatter

\section*{List of abbreviations}
API: application programming interface;
CI: confidence interval;
GPU: graphics processing unit;
IPIP-NEO-120: 120-item International Personality Item Pool representation of the
NEO Personality Inventory;
LLM: large language model;
OLS: ordinary least squares;
SD: standard deviation.

\bmhead{Supplementary information}
Additional file 1 contains the complete prompt texts, the full condition-by-metric results,
every contrast tested with effect sizes and bootstrap intervals, the per-perturbation values
of the robustness audit, the collective-intelligence item-screening record, and the
validity-principle compliance table.

\section*{Declarations}

\bmhead{Availability of data and materials}
The datasets generated and analysed during the current study, comprising the run-level
results for all 991 simulations together with the derived analysis tables, the induction
validation record, the collective-intelligence item-screening record and the reconstructed
realised trait distributions, are available in the Zenodo repository,
\url{https://doi.org/10.5281/zenodo.21792633}, and in the project repository,
\url{https://github.com/raadbintareaf/traitmix}. All input data are public: the
IPIP-NEO-120 items are in the public domain (\url{https://ipip.ori.org}); ground-truth
quantities for the estimation tasks were retrieved from the World Bank Open Data API
(\url{https://data.worldbank.org}); normative personality statistics are taken from the
published open-access source cited in the text.

\noindent The simulation framework and analysis pipeline are released as follows.
Project name: TraitMix.
Project home page: \url{https://github.com/raadbintareaf/traitmix}.
Archived version: \url{https://doi.org/10.5281/zenodo.21792633}.
Operating system: platform independent; developed and run on Linux.
Programming language: Python 3.12.
Other requirements: vLLM 0.8.5.post1 and PyTorch 2.6.0 with CUDA 12.4 for the simulation
runs, which require a GPU with at least 24\,GB of memory; the analysis and figure pipeline
requires only a CPU.
License: MIT.
Restrictions to use by non-academics: none.

\bmhead{Competing interests}
The author declares that he has no competing interests.

\bmhead{Funding}
This work was supported by a research grant from the German University of Digital Science.
The funder had no role in the conceptualization, design, data collection, analysis, decision
to publish, or preparation of the manuscript.

\bmhead{Authors' contributions}
RBT is the sole author. RBT conceived the study, designed the methodology, implemented the
simulation framework and analysis pipeline, conducted the experiments, curated the data,
produced the visualisations, and wrote the manuscript. The author read and approved the
final manuscript.

\bmhead{Acknowledgements}
Not applicable.

\bibliography{sn-bibliography}

\clearpage
\includepdf[pages=-]{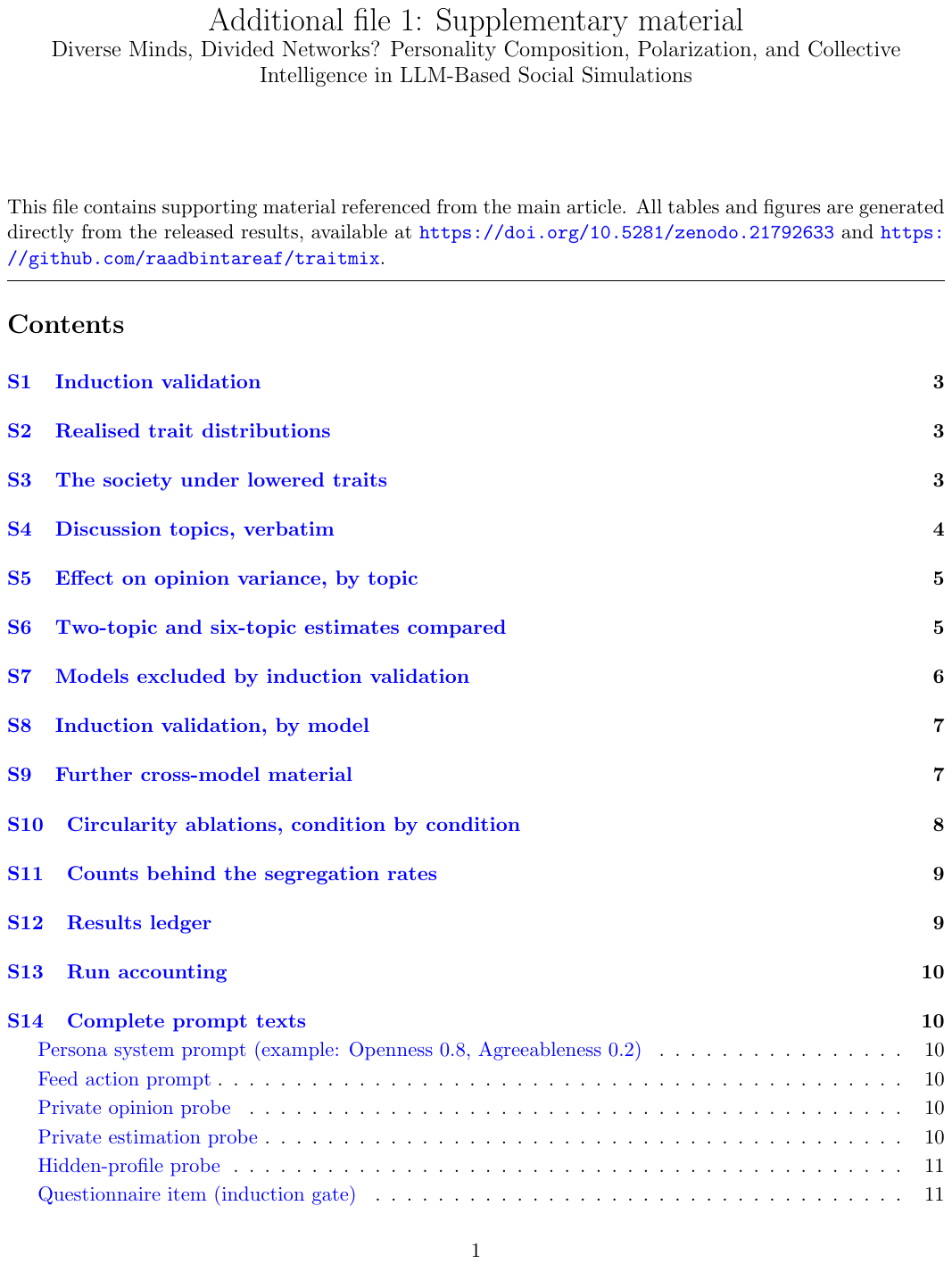}

\end{document}